\documentclass[trackchanges,twocolumn,twocolappendix]{aastex701}

\usepackage{hyperref}
\usepackage{appendix}
\usepackage{amsmath}
\usepackage{amssymb}
\usepackage{mathtools}

\makeatletter
\renewcommand{\maketag@@@}[1]{\hbox{\normalsize#1}}
\makeatother

\makeatletter
\def\tablecomments#1{\vskip1pt{\small\vskip1sp\noindent\vrule height 11pt depth 2pt
		width 0pt\currtabletypesize\parindent=0pt\leftskip=0pt #1\vskip1pt}}
\makeatother

\makeatletter
\newcommand\thefontsize[1]{{#1 The current font size is: \f@size pt\par}}
\makeatother

\shorttitle{New Candidate Activity Periods from Repeating~FRB Sources}

\shortauthors{Pearlman et al.}

\graphicspath{{./}{figures/}}

\begin{document}

\title{New Candidate Activity Periods from FRB~20240209A and \\ Repeating Fast Radio Burst Sources in the Second \mbox{CHIME/FRB}~Catalog}

\author[orcid=0000-0002-8912-0732]{Aaron~B.~Pearlman}
\email{aaron.b.pearlman@mit.edu}
\correspondingauthor{\href{mailto:aaron.b.pearlman@mit.edu}{aaron.b.pearlman@mit.edu}}
\altaffiliation{NASA Hubble Fellow.}
\affiliation{MIT Kavli Institute for Astrophysics and Space Research, Massachusetts Institute of Technology, 77 Massachusetts Avenue, Cambridge, MA 02139, USA}
\affiliation{Department of Physics, McGill University, 3600 rue University, Montr\'eal, QC H3A 2T8, Canada}
\affiliation{Trottier Space Institute, McGill University, 3550 rue University, Montr\'eal, QC H3A 2A7, Canada}

\author[orcid=0009-0006-1258-4228]{Maxwell~Fine}
\email{maxwell.fine@mail.mcgill.ca}
\affiliation{Department of Physics, McGill University, 3600 rue University, Montr\'eal, QC H3A 2T8, Canada}
\affiliation{Trottier Space Institute, McGill University, 3550 rue University, Montr\'eal, QC H3A 2A7, Canada}

\author[orcid=0000-0001-6422-8125]{Amanda~M.~Cook}
\email{amanda.cook@mail.mcgill.ca}
\affiliation{Department of Physics, McGill University, 3600 rue University, Montr\'eal, QC H3A 2T8, Canada}
\affiliation{Trottier Space Institute, McGill University, 3550 rue University, Montr\'eal, QC H3A 2A7, Canada}
\affiliation{Anton Pannekoek Institute for Astronomy, University of Amsterdam, Science Park 904, 1098 XH Amsterdam, The Netherlands}

\author[orcid=0000-0001-9345-0307]{Victoria~M.~Kaspi}
\email{victoria.kaspi@mcgill.ca}
\affiliation{Department of Physics, McGill University, 3600 rue University, Montr\'eal, QC H3A 2T8, Canada}
\affiliation{Trottier Space Institute, McGill University, 3550 rue University, Montr\'eal, QC H3A 2A7, Canada}
\affiliation{School of Physics and Astronomy, Tel Aviv University, Tel Aviv 69978, Israel}

\author[orcid=0000-0002-4279-6946]{Kiyoshi~W.~Masui}
\email{kmasui@mit.edu}
\affiliation{MIT Kavli Institute for Astrophysics and Space Research, Massachusetts Institute of Technology, 77 Massachusetts Avenue, Cambridge, MA 02139, USA}
\affiliation{Department of Physics, Massachusetts Institute of Technology, 77 Massachusetts Avenue, Cambridge, MA 02139, USA}

\author[orcid=0009-0005-5370-7653]{Daniel~Amouyal}
\email{daniel.amouyal@mail.mcgill.ca}
\affiliation{Department of Physics, McGill University, 3600 rue University, Montr\'eal, QC H3A 2T8, Canada}
\affiliation{Trottier Space Institute, McGill University, 3550 rue University, Montr\'eal, QC H3A 2A7, Canada}

\author[orcid=0000-0002-2878-1502]{Shami~Chatterjee}
\email{shami@astro.cornell.edu}
\affiliation{Cornell Center for Astrophysics and Planetary Science, Cornell University, Ithaca, NY 14853, USA}

\author[orcid=0000-0002-8376-1563]{Alice~P.~Curtin}
\email{alice.curtin@mcgill.ca}
\affiliation{Department of Physics, McGill University, 3600 rue University, Montr\'eal, QC H3A 2T8, Canada}
\affiliation{Trottier Space Institute, McGill University, 3550 rue University, Montr\'eal, QC H3A 2A7, Canada}
\affiliation{Anton Pannekoek Institute for Astronomy, University of Amsterdam, Science Park 904, 1098 XH Amsterdam, The Netherlands}

\author[orcid=0000-0001-7166-6422]{Matt~Dobbs}
\email{matt.dobbs@mcgill.ca}
\affiliation{Department of Physics, McGill University, 3600 rue University, Montr\'eal, QC H3A 2T8, Canada}
\affiliation{Trottier Space Institute, McGill University, 3550 rue University, Montr\'eal, QC H3A 2A7, Canada}

\author[orcid=0000-0003-0307-9984]{Tarraneh~Eftekhari}
\email{teftekhari@northwestern.edu}
\affiliation{Center for Interdisciplinary Exploration and Research in Astrophysics, Northwestern University, 1800 Sherman Avenue, Evanston, IL 60201, USA}

\author[orcid=0000-0001-8384-5049]{Emmanuel~Fonseca}
\email{emmanuel.fonseca@mail.wvu.edu}
\affiliation{Department of Physics and Astronomy, West Virginia University, PO Box 6315, Morgantown, WV 26506, USA}
\affiliation{Center for Gravitational Waves and Cosmology, West Virginia University, Chestnut Ridge Research Building, Morgantown, WV 26505, USA}

\author[orcid=0000-0002-3382-9558]{B.~M.~Gaensler}
\email{gaensler@ucsc.edu}
\affiliation{Department of Astronomy and Astrophysics, University of California, Santa Cruz, 1156 High Street, Santa Cruz, CA 95064, USA}
\affiliation{Dunlap Institute for Astronomy and Astrophysics, University of Toronto, 50 St. George Street, Toronto, ON M5S 3H4, Canada}
\affiliation{David A. Dunlap Department of Astronomy and Astrophysics, University of Toronto, 50 St. George Street, Toronto, ON M5S 3H4, Canada}

\author[orcid=0000-0003-2317-1446]{Jason~W.~T.~Hessels}
\email{jason.hessels@mcgill.ca}
\affiliation{Department of Physics, McGill University, 3600 rue University, Montr\'eal, QC H3A 2T8, Canada}
\affiliation{Trottier Space Institute, McGill University, 3550 rue University, Montr\'eal, QC H3A 2A7, Canada}
\affiliation{Anton Pannekoek Institute for Astronomy, University of Amsterdam, Science Park 904, 1098 XH Amsterdam, The Netherlands}
\affiliation{ASTRON, Netherlands Institute for Radio Astronomy, Oude Hoogeveensedijk 4, 7991 PD Dwingeloo, The Netherlands}

\author[orcid=0009-0004-4176-0062]{Afrasiyab~(Afrokk)~Khan}
\email{afrasiyab.khan@mcgill.ca}
\affiliation{Department of Physics, McGill University, 3600 rue University, Montr\'eal, QC H3A 2T8, Canada}
\affiliation{Trottier Space Institute, McGill University, 3550 rue University, Montr\'eal, QC H3A 2A7, Canada}

\author[orcid=0000-0001-7931-0607]{Dongzi~Li}
\email{dzli@tsinghua.edu.cn}
\affiliation{Department of Astronomy, Tsinghua University, Beijing 100084, China}

\author[orcid=0000-0001-7348-6900]{Ryan~Mckinven}
\email{ryan.mckinven@mcgill.ca}
\affiliation{Department of Physics, McGill University, 3600 rue University, Montr\'eal, QC H3A 2T8, Canada}
\affiliation{Trottier Space Institute, McGill University, 3550 rue University, Montr\'eal, QC H3A 2A7, Canada}

\author[orcid=0000-0002-8897-1973]{Ayush~Pandhi}
\email{ayush.pandhi@mcgill.ca}
\affiliation{Department of Physics, McGill University, 3600 rue University, Montr\'eal, QC H3A 2T8, Canada}
\affiliation{Trottier Space Institute, McGill University, 3550 rue University, Montr\'eal, QC H3A 2A7, Canada}

\author[orcid=0000-0002-4795-697X]{Ziggy~Pleunis}
\email{z.pleunis@uva.nl}
\affiliation{Anton Pannekoek Institute for Astronomy, University of Amsterdam, Science Park 904, 1098 XH Amsterdam, The Netherlands}
\affiliation{ASTRON, Netherlands Institute for Radio Astronomy, Oude Hoogeveensedijk 4, 7991 PD Dwingeloo, The Netherlands}

\author[orcid=0000-0002-4623-5329]{Mawson~W.~Sammons}
\email{mawson.sammons@lam.fr}
\affiliation{Laboratoire d'Astrophysique de Marseille, Aix-Marseille Universit\'e, CNRS, CNES, Marseille, France}

\author[orcid=0000-0003-3154-3676]{Ketan~R.~Sand}
\email{ketan.sand@mail.mcgill.ca}
\affiliation{Department of Physics, McGill University, 3600 rue University, Montr\'eal, QC H3A 2T8, Canada}
\affiliation{Trottier Space Institute, McGill University, 3550 rue University, Montr\'eal, QC H3A 2A7, Canada}

\author[orcid=0000-0002-7374-7119]{Paul~Scholz}
\email{pscholz@yorku.ca}
\affiliation{Department of Physics and Astronomy, York University, 4700 Keele Street, Toronto, ON M3J 1P3, Canada}

\author[orcid=0000-0002-4823-1946]{Vishwangi~Shah}
\email{vishwangi.shah@mail.mcgill.ca}
\affiliation{Department of Physics, McGill University, 3600 rue University, Montr\'eal, QC H3A 2T8, Canada}
\affiliation{Trottier Space Institute, McGill University, 3550 rue University, Montr\'eal, QC H3A 2A7, Canada}

\author[orcid=0000-0002-6823-2073]{Kaitlyn~Shin}
\email{kaitshin@caltech.edu}
\affiliation{Cahill Center for Astronomy and Astrophysics, California Institute of Technology, MC 249-17, 1200 E. California Boulevard, Pasadena, CA 91125, USA}

\author[orcid=0000-0002-2088-3125]{Kendrick~Smith}
\email{kmsmith@perimeterinstitute.ca}
\affiliation{Perimeter Institute for Theoretical Physics, 31 Caroline Street North, Waterloo, ON N2L 2Y5, Canada}

\author[orcid=0000-0001-9784-8670]{Ingrid~H.~Stairs}
\email{stairs@astro.ubc.ca}
\affiliation{Department of Physics and Astronomy, University of British Columbia, 6224 Agricultural Road, Vancouver, BC V6T 1Z1, Canada}


\begin{abstract}

Repeating fast radio burst~(FRB) sources with periodic activity provide important constraints on FRB~models invoking neutron-star precession, rotation, or orbital modulation in binary systems. Although a few FRB~sources exhibit evidence of periodic behavior, it remains unclear whether periodic activity is a common underlying feature of repeating~FRBs. Using a new FRB periodicity search pipeline introduced by~\citet{Pearlman+2026a}, which is optimized for sparse event-arrival-time data from FRB~surveys, we search for activity periods in the burst arrival times from FRB~20240209A and repeating~FRBs in the Second Canadian Hydrogen Intensity Mapping Experiment~\mbox{(CHIME)/FRB} Catalog. For FRB~20240209A, we identify a new candidate activity period of 170.4\,$\pm$\,4.7\,days, with a trials-corrected statistical significance of 5.3$\sigma$. Our larger burst sample shows no evidence of the \mbox{128.5-day} activity period previously reported for this source by~\citet{Pal2025}. We also report candidate activity periods of 28.46\,$\pm$\,0.08\,days, 86.7\,$\pm$\,0.5\,days, and 36.4\,$\pm$\,0.5\,days for FRB~20190208A, FRB~20200926A, and FRB~20220529A, with trials-corrected statistical significances of 3.1$\sigma$, 4.0$\sigma$, and 3.5$\sigma$, respectively. All statistical significances are evaluated against a homogeneous (constant-rate) Poisson null hypothesis, and we caution that they may decrease under null models that incorporate burst clustering. Collectively, our results constitute the most comprehensive search for periodic activity in the repeating~FRB population to date.

\end{abstract}


\keywords{\uat{Radio transient sources}{2008}; \uat{Radio bursts}{1339}; \uat{High energy astrophysics}{739};  \uat{Radio telescopes}{1360}; \uat{Transient sources}{1851}; \uat{Burst astrophysics}{187}}


\section{Introduction}
\label{sec:introduction}

It has been nearly two decades since the discovery of the first fast radio burst~(FRB;~\citealt{Lorimer+2007}) and a decade since the first repeating~FRB source was identified~\citep{Spitler+2016}, yet the physical nature of these extragalactic radio transients remains a major open question. The detection of an FRB-like burst from the Galactic magnetar SGR~1935$+$2154~\citep{CHIME+2020a, Bochenek+2020c} was a key breakthrough, providing direct evidence that magnetars can produce FRB-like emission. However, it is still unclear whether magnetars power most~FRBs, or whether~FRBs are produced by multiple types of sources~\citep{Platts+2019, Petroff+2019, Cordes+2019, Bhardwaj+2021a, Pleunis+2021a, Kirsten+2022, Petroff+2022, Bailes2022, Zhang2023, Pearlman+2025a}.

The discovery of the \mbox{16.3-day} periodic activity from the repeating FRB~20180916B was an important milestone in the study of~FRBs~\citep{CHIME+2020b}. The source exhibits regular, predictable bursting during a \mbox{4--5-day} activity window each cycle in the \mbox{CHIME/FRB} band~(\mbox{400--800}\,MHz;~\citealt{CHIME+2020b, Sand+2023, Pearlman+2026a}). FRB~20180916B's activity window is also chromatic, occurring earlier and becoming narrower at higher radio frequencies~\citep{Pleunis+2021b, Pastor-Marazuela+2021, Bethapudi+2023, Gopinath+2024}. However, neither its burst morphology (e.g.,~scattering timescale and burst width) nor its propagation properties (e.g.,~dispersion measure~(DM) and Faraday rotation measure~(RM)) show variations that track the \mbox{16.3-day} activity period~\citep{Mckinven+2023a, Sand+2023, Bethapudi+2025, Sand+2025}.

Candidate periodicities have also been proposed for other repeaters, though it remains unclear whether these arise from the same mechanism responsible for FRB~20180916B's well-established periodicity. The first repeating~FRB source identified, FRB~20121102A, exhibits tentative evidence of a \mbox{$\sim$160-day} periodicity in its burst activity, with a relatively large duty cycle of~$\sim$60\% (i.e., an activity window of $\sim$100\,days;~\citealt{Rajwade+2020b, Cruces+2021, Braga+2025, Gouiffes+2025, Wang+2025}). \citet{Li+2024a} further suggest the presence of an additional \mbox{4.6-day} periodicity from FRB~20121102A, primarily associated with its most energetic bursts. More recently, a \mbox{128.5-day} activity period has been reported for the hyperactive repeating FRB~20240209A~\citep{Pal2025}, although the data analyzed in~\citet{Pal2025} span only approximately three putative activity cycles. Candidate periodic signals on long timescales have also been reported in the temporal variations of the burst~RMs of some repeaters, including FRB~20201124A~\citep{Xu+2025} and FRB~20220529A~\citep{Liang+2025}.

Although a variety of models have been proposed to explain the days- to months-long periodicities observed in a small fraction of repeating~FRBs, their origin remains uncertain. Precession models invoke a highly magnetized neutron star whose strong internal magnetic field deforms the star, with the resulting asphericity leading to free precession~(e.g.,~\citealt{Levin+2020, Zanazzi+2020, Wasserman+2022}). \citet{Beniamini+2020a} instead argue that the periodicity reflects the spin
period of a magnetar spun down to ultra-long periods by mass-loaded particle winds, fallback accretion, or both. Orbital models attribute the periodic activity to an FRB~source in a binary system, in which bursts are preferentially observed at certain orbital phases~(e.g.,~\citealt{Lyutikov+2020a, Deng+2021, Sridhar+2021}). In the ``binary comb'' model, for example, a strong wind from a massive stellar or millisecond pulsar companion combs the magnetosphere of the FRB~source into a funnel pointing away from the companion, through which bursts can escape. The funnel sweeps across our line of sight once per orbit, producing the periodic activity window~\citep{Ioka+2020}. Related models invoking eccentric orbits with a planetary companion~\citep{Kurban+2022} or interactions with asteroid belts~\citep{Du+2021} have also been explored.

Alternatively, the observed periodicity may arise from a source that is active for a characteristic duration and then turns off while it replenishes the energy reservoir powering its bursts, producing quasi-regular activity cycles. Such behavior can mimic periodic activity, particularly when only a few cycles are sampled~\citep{Vaughan+2016}. One potential way to distinguish between these scenarios is to measure a period derivative~\citep{Katz+2021}. However, no significant period derivative has yet been measured for any periodic repeater (e.g.,~\citealt{Sand+2023, Pearlman+2026a} for FRB~20180916B). Increasing the sample of periodic repeaters, and monitoring them over many activity cycles, is therefore essential for testing these models.

Here, we present a systematic search for periodicities on timescales of days to years in the burst arrival times from repeating~FRBs in the Second \mbox{CHIME/FRB} Catalog (hereafter \mbox{CHIME/FRB} Catalog~2;~\citealt{CHIME+2026}), using the FRB~periodicity search pipeline developed by~\citet{Pearlman+2026a} and taking advantage of the near-daily exposure provided by \mbox{CHIME/FRB} over a multi-year baseline. \mbox{CHIME/FRB} Catalog~2 includes both new and previously discovered repeaters~\citep{CHIME+2019a, CHIME+2019b, Fonseca+2020, CHIME+2021b, Bhardwaj+2021a, Bhardwaj+2021b, Lanman+2022, CHIME+2023a, Shah+2025, Shin+2026, CHIME+2026, Abbott+2026, Cook+2026}, and it contains the largest sample of repeating~FRB sources detected with a single instrument~\citep{Cook+2026}, allowing us to assess how common periodic activity is among repeating~FRBs. We also test the \mbox{128.5-day} activity period reported by~\citet{Pal2025} for FRB~20240209A using more bursts over a longer baseline. This source is not included in \mbox{CHIME/FRB} Catalog~2 because it was discovered after the catalog observing period.

This paper is organized as follows. In Section~\ref{sec:observations}, we describe the \mbox{CHIME/FRB}~observations and introduce the FRB~data sets analyzed in this work. In Section~\ref{sec:frb_periodicity_search}, we provide an overview of the FRB~periodicity search pipeline introduced in~\citet{Pearlman+2026a} and describe the heuristic criterion used to define the sample of repeating~FRBs included in our analysis. In Section~\ref{sec:results}, we present the results of our searches for periodic activity in FRB~20240209A and the repeating~FRBs in \mbox{CHIME/FRB} Catalog~2. In Section~\ref{sec:discussion}, we discuss the astrophysical implications of our results. Finally, in Section~\ref{sec:conclusions}, we summarize our main conclusions and discuss the prospects for future discoveries.


\section{\mbox{CHIME/FRB}~Observations and Data}
\label{sec:observations}


\subsection{The \mbox{CHIME/FRB}~Instrument and FRB~Data}
\label{sec:chime_frb_data}

The burst arrival times used in this work were obtained with CHIME~\citep{CHIME+2022b}, a transit radio telescope located at the Dominion Radio Astrophysical Observatory in Penticton, British Columbia, Canada. CHIME consists of four stationary cylindrical reflectors, each oriented North--South, and operates over the \mbox{400--800}\,MHz band. Since the telescope has no moving parts, it observes the entire sky north of decl.\,$\approx$\,\mbox{$-$9.5$^{\circ}$} once per sidereal day as the Earth rotates. Sources at decl.\,$\gtrsim$\,\mbox{$+$70$^{\circ}$} are observed twice per sidereal day, during their upper and lower transits, with different sensitivities and effective exposures in the two transits~\citep{CHIME+2021b, CHIME+2026}.

The \mbox{CHIME/FRB} backend performs a real-time search for dispersed, millisecond-duration bursts in 1024~formed beams that tile CHIME's primary field of view~\citep{CHIME+2018}. Before the FRB~search, the intensity data are \mbox{up-channelized} to a frequency resolution of 24.4\,kHz and downsampled to a time resolution of 0.983\,ms. For each detected candidate, the total-intensity dynamic spectrum is saved for offline analysis, which enables detailed burst modeling and precise measurement of the burst time of arrival~(ToA).


\subsection{FRB~20240209A}
\label{sec:obs_frb20240209a}

FRB~20240209A is a repeating~FRB source localized to the outskirts of a massive, quiescent elliptical galaxy~\citep{Eftekhari+2025, Shah+2025}. This environment differs markedly from the star-forming hosts typically associated with formation channels involving young neutron stars (e.g.,~\citealt{Bassa+2017, Marcote+2020, Heintz+2020, Bhandari+2022, Gordon+2023}). Sources in such environments broaden the range of plausible FRB~progenitor pathways and highlight the importance of studying repeaters across diverse host environments.

FRB~20240209A has also been reported to exhibit periodic activity with a period of 128.5\,days~\citep{Pal2025}, making it a particularly compelling target for a dedicated analysis. Since FRB~20240209A was discovered after the end of the \mbox{CHIME/FRB} Catalog~2 observing period, it is not included in the catalog. We therefore analyze it as a single-source data set, separate from the catalog repeater sample described in Section~\ref{sec:obs_chime_frb_cat2}.

We compiled all \mbox{CHIME/FRB}~detections of FRB~20240209A between 2018~July~25 and 2025~December~31, using a combination of automated and manual methods. Candidate bursts were first identified using a \texttt{DBSCAN}-based clustering algorithm~\citep{Ester+1996, Dong+2023, CHIME+2023a, Cook+2026}, which compares the DM and header-localized position~\citep{CHIME+2021b, CHIME+2026} of each detected event to the known DM and VLBI~position of the source~\citep{Shah+2025}. We then computed an association probability~($P_{\mathrm{assoc}}$) for each candidate, following the procedure described in Appendix~C of~\citet{Pearlman+2026a}, and retained those with $P_{\mathrm{assoc}}$\,$\geq$\,0.5 for further verification. We chose this threshold, which is lower than the threshold that is typically adopted for other repeating sources, to maximize the completeness of the burst sample. To preserve its purity, we visually inspected the header localization and total-intensity data of every retained candidate to confirm the presence of FRB~emission and a confident association with FRB~20240209A. Both steps were necessary because FRB~20240209A lies near the north celestial pole~(decl.\,$\approx$\,$+$86$^{\circ}$), where source association is complicated by geometric projection effects and by the high \mbox{CHIME/FRB} detection density in this region of the sky.

This procedure yielded 155~bursts from FRB~20240209A, including those reported in~\citet{Shah+2025}. For consistency, all burst arrival times were measured directly from \mbox{CHIME/FRB} total-intensity~(Stokes~I) data, sampled at 0.983\,ms resolution. The~ToA, signal-to-noise ratio~\mbox{(S/N)}, and DM of each burst are listed in Table~\ref{table:frb20240209a_burst_log}, and a timeline of the burst~ToAs is shown in Figure~\ref{fig:frb_toa_timeline}.


\subsection{Repeating FRBs in \mbox{CHIME/FRB} Catalog~2}
\label{sec:obs_chime_frb_cat2}

Our second data set consists of burst arrival times from repeating~FRB sources in \mbox{CHIME/FRB} Catalog~2~\citep{CHIME+2026, Cook+2026}, a uniformly processed sample of~FRBs detected between 2018~July~25 and 2023~September~15~(UTC, inclusive). The catalog contains 4539~bursts from 3641~unique sources, including 981~bursts from 83~repeating sources. These repeaters were identified through a combination of real-time association with previously known sources and offline clustering in \mbox{position--DM} space. We adopt the resulting repeater identifications in~\citet{Cook+2026}, which also presents a complementary analysis of the population statistics of these sources. The catalog's five-year timing baseline and large repeater sample make it particularly well suited to searches for periodicities in repeater burst arrival times.

\begin{figure*}
	\centering
	\includegraphics[height=0.90\textheight, trim=0bp 15bp 0bp 0bp]{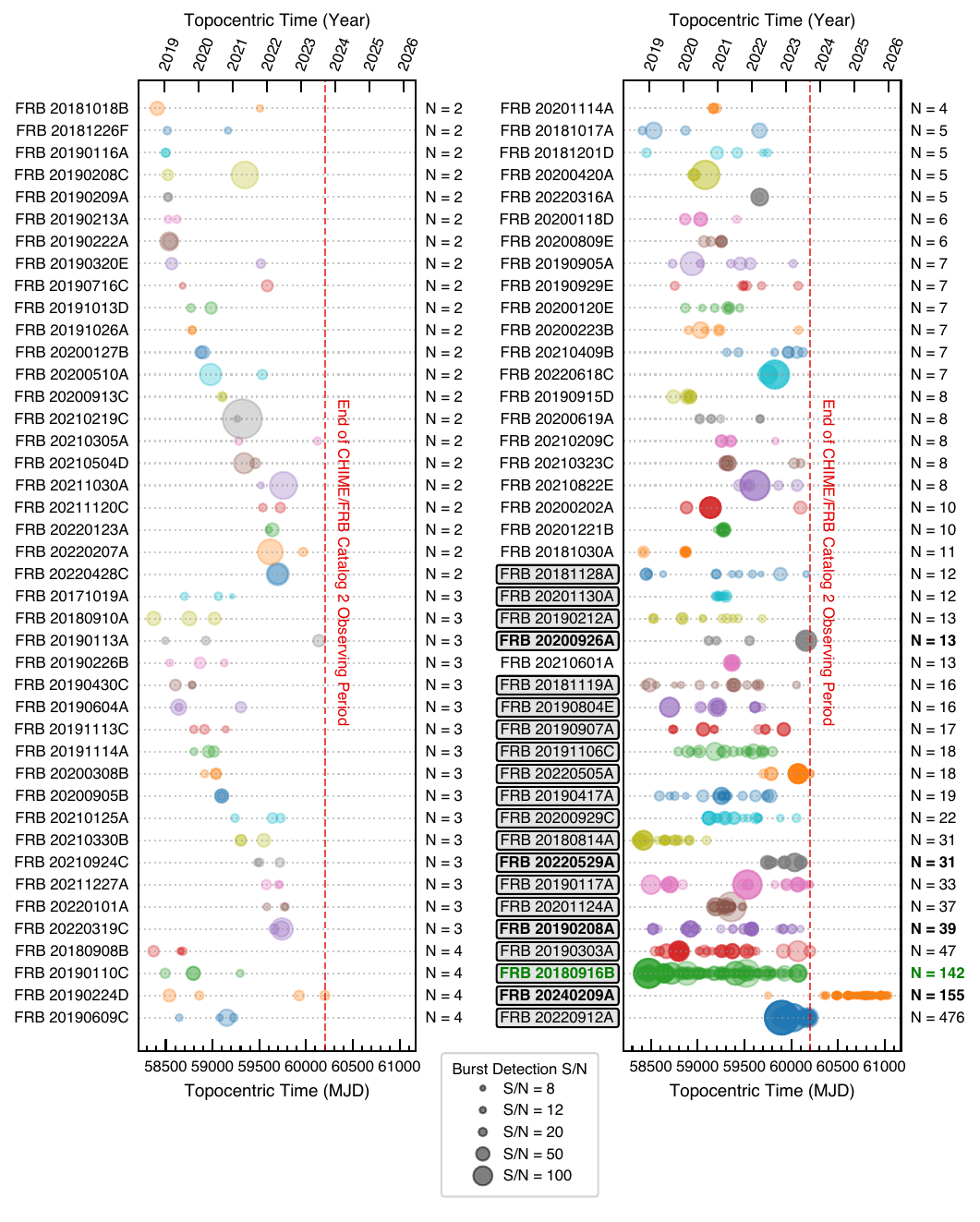}
	\caption{Timeline of burst arrival times from the repeating~FRB sources considered in this work. Each circle represents a detected burst, with its size scaled by the detection~\mbox{S/N}. The Transient Name Server~(TNS; see~\url{https://www.wis-tns.org}) name of each repeater is shown on the left in each column, and the number of burst~ToAs detected with \mbox{CHIME/FRB}~is labeled on the right in each column. The sources are ordered by increasing number of~ToAs, and sources with an equal number of~ToAs are ordered by their TNS~name (i.e.,~chronologically by discovery date). The TNS~names of sources included in our periodicity search are enclosed in gray rectangles, and those with a confirmed or candidate activity period are highlighted in bold. FRB~20180916B, which has a well-established \mbox{16.3-day} activity period~\citep{CHIME+2020b, Pleunis+2021b, Pastor-Marazuela+2021, Sand+2023, Pearlman+2026a}, is highlighted in green. The vertical dashed red line marks the end of the \mbox{CHIME/FRB} Catalog~2 observing period. FRB~20121102A is not included in this sample because it has been detected only once with \mbox{CHIME/FRB}~\citep{Josephy+2019}.}
	\label{fig:frb_toa_timeline}
\end{figure*}

For each event, \mbox{CHIME/FRB} Catalog~2\footnote{The \mbox{CHIME/FRB} Catalog~2 data are publicly available at \href{https://www.chime-frb.ca/catalog2}{https://www.chime-frb.ca/catalog2}.} provides burst metadata, sky localizations, and survey characterization products, of which the burst arrival times, repeater associations, and exposure estimates are most relevant to this work. The FRB~ToAs are reported as topocentric~MJDs, referenced both to 400.1953125\,MHz~(\texttt{mjd\_400}) and to infinite frequency~(\texttt{mjd\_inf}), with associated uncertainties (\texttt{mjd\_400\_err} and \texttt{mjd\_inf\_err}, respectively).\footnote{A descriptive list of measured parameters in \mbox{CHIME/FRB}~Catalog~2 is available at \href{https://www.chime-frb.ca/information?tab=catalog2-glossary}{https://www.chime-frb.ca/information?tab=catalog2-glossary}.} We adopt the cataloged \texttt{mjd\_400} values as our input~ToAs and group bursts into source-specific event lists using the repeater associations encoded in the \texttt{repeater\_name} field. We treat each cataloged burst~ToA as an independent event in our periodicity analysis. The repeating~FRB sources considered here are shown in Figure~\ref{fig:frb_toa_timeline}, together with a timeline of their burst arrival times.


\subsection{\mbox{CHIME/FRB}~Source Exposure}
\label{sec:chime_frb_exposure}

Although \mbox{CHIME/FRB} observes every source in its field of view at least once per sidereal day, its effective exposure is non-uniform because of telescope downtime, time-dependent beam availability, sensitivity fluctuations, and the declination-dependent transit geometry of the instrument. We therefore construct an exposure function for each repeater considered in this work, following the framework described in~\citet{CHIME+2021b}. These exposure functions define the observational window function of each source, which we incorporate into the significance estimates in our FRB~periodicity search~(see Section~\ref{sec:frb_periodicity_search_pipeline_overview}).

\begin{figure}
	\centering
	\includegraphics[width=\linewidth, trim=6bp 6bp 6bp 6bp, clip]{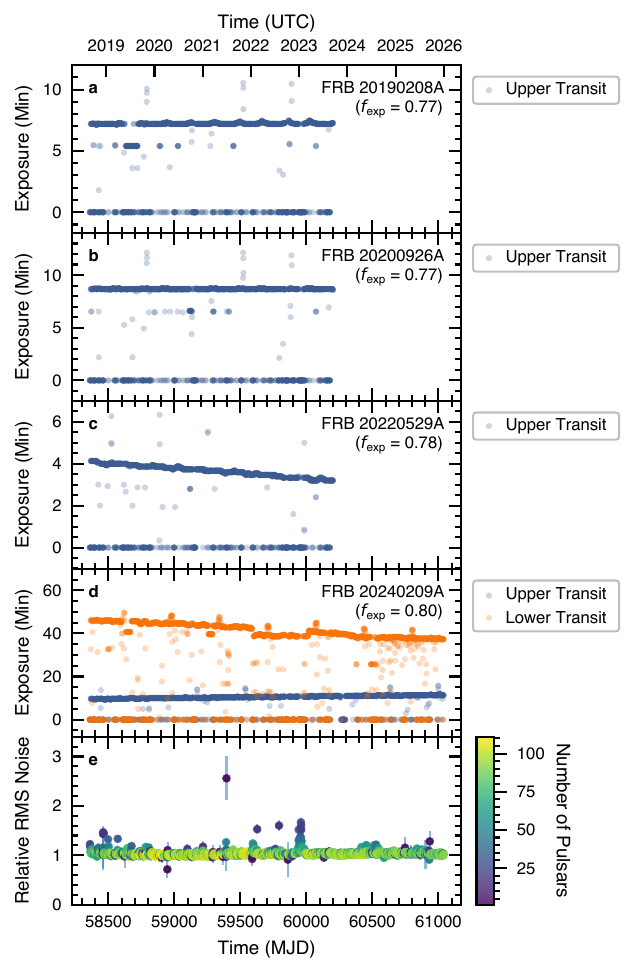}
	\caption{Daily \mbox{CHIME/FRB}~exposure for the four repeating~FRB sources with candidate periodicities identified in this work, shown together with the relative RMS~noise. Panels~\mbox{a--c} show the daily exposure of FRB~20190208A, FRB~20200926A, and FRB~20220529A, respectively, from the end of the pre-commissioning period to the end of the \mbox{CHIME/FRB} Catalog~2 observing period~\mbox{(2018~August~28--2023~September~15}~UTC, inclusive). Panel~d shows the daily exposure of FRB~20240209A between 2018~August~28 and 2025~December~31~(UTC, inclusive). The fraction of daily transits with non-zero exposure~($f_{\mathrm{exp}}$) is labeled in the top-right corner of panels~\mbox{a--d}. Panel~e shows the relative RMS~noise over the same time interval as panel~d, derived from the \mbox{S/N}~distributions of Galactic radio pulsars, with the color indicating the number of pulsars used to estimate each value. Since the sidereal day is shorter than the solar day, the transit time drifts through the UTC~day over the course of a year, and two transits occasionally fall within the same UTC~day, approximately doubling the nominal daily exposure. Brief, discrete drops in the exposure~(e.g.,~panels~a and~b) occur when a compute node processing data from one of the beams through which the source transits is offline~\citep{CHIME+2018}. The gradual decline in some exposures (e.g.,~panels~c and~d) arises because precession of Earth's rotation axis slowly shifts the apparent source position relative to the \mbox{CHIME/FRB}~formed beams.}
	\label{fig:frb_source_exposure}
\end{figure}

We compute the daily exposure of each source using its most precise available sky position. These positions are obtained from \mbox{CHIME/FRB} baseband localizations~\citep{Michilli+2021, Cook+2026}, very-long-baseline interferometry~(VLBI) localizations with \mbox{CHIME/FRB} and its Outrigger stations~\citep{CHIME+2025c, Shah+2025}, or VLBI~localizations with the European VLBI Network~\citep{Nimmo+2022b, Hewitt+2024a, Moroianu+2026}. The daily exposure is defined as the total time that the source position lies within the full width at half maximum~(FWHM) of any active formed beam at 600\,MHz. Depending on the source declination, exposure may accumulate during the upper transit, the lower transit, or both, and we track these contributions separately. We exclude beams whose designated compute node was offline, since no data from those beams were processed. We also exclude transits during the pre-commissioning period \mbox{(2018~July~25--August~27)}, when \mbox{CHIME/FRB} was operating with a different beam configuration and its sensitivity and exposure were not yet well characterized. \mbox{Figure~\ref{fig:frb_source_exposure}a--d} shows the daily exposure for the four repeating FRBs with candidate periodicities identified in this work: FRB~20190208A, FRB~20200926A, FRB~20220529A, and FRB~20240209A.

To monitor time-dependent sensitivity variations, we use daily measurements of the relative root-mean-square~(RMS) noise, derived from the \mbox{S/N}~distributions of Galactic radio pulsars that \mbox{CHIME/FRB} uses for calibration~\citep{Josephy+2019, CHIME+2019b, Fonseca+2020, CHIME+2021b, CHIME+2026}. These measurements trace day-to-day variations in the effective system sensitivity (e.g.,~from changes in the gain calibration, weather, and radio-frequency interference). We exclude from the exposure calculation all days on which the RMS~noise exceeds its baseline value by more than 10\%. For the sources analyzed here, the RMS~variations are modest and do not qualitatively affect the exposure model. Figure~\ref{fig:frb_source_exposure}e shows the daily RMS noise, with the color indicating the number of pulsars used to estimate each value.


\section{FRB~Periodicity Search}
\label{sec:frb_periodicity_search}

In this section, we describe our search for periodicities in the burst arrival times from repeating~FRB sources. We first provide an overview of our FRB~periodicity search pipeline (see Section~\ref{sec:frb_periodicity_search_pipeline_overview}), which is presented in detail in a companion paper~\citep{Pearlman+2026a}. We then introduce a heuristic criterion for selecting the subset of repeaters in \mbox{CHIME/FRB} Catalog~2 for which a periodicity search is likely to be informative (see Section~\ref{sec:cat2_frb_source_selection}).


\subsection{FRB~Periodicity Search Pipeline Overview}
\label{sec:frb_periodicity_search_pipeline_overview}

We search each set of FRB~ToAs for periodicities using the pipeline described in~\citet{Pearlman+2026a}. The trial-frequency range for each source is
\begin{equation}
	\left[f_{\mathrm{min}}, f_{\mathrm{max}}\right] = \left[\frac{3}{T}, \frac{1}{P_{\mathrm{min}}}\right],
	\label{eqn:trial_freq_range}
\end{equation}
where $T$\,$=$\,$t_{\mathrm{max}}-t_{\mathrm{min}}$ is the time span of the burst~ToAs, $t_{\mathrm{min}}$ and $t_{\mathrm{max}}$ are the first and last arrival times in the ToA~set, and $P_{\mathrm{min}}$ is the minimum trial period (see Section~2.2 in~\citealt{Pearlman+2026a}). The lower bound ensures that the arrival times span at least three cycles of any candidate periodicity. We adopt a minimum trial period of $P_{\mathrm{min}}$\,$=$\,10\,days to restrict the search to long-timescale periodic signals and to mitigate sidereal aliasing from \mbox{CHIME/FRB}'s daily observing cadence. We oversample the frequency grid by a factor of $O$\,$=$\,5 relative to the intrinsic Fourier resolution~($\Delta f$\,$=$\,$1/T$). The range of trial periods searched for each source is listed in Table~\ref{table:frb_periodicity_search_summary}.

For each source, we compute three complementary periodograms on this frequency grid: the \mbox{Lomb--Scargle}~(LS) periodogram~\citep{Lomb1976, Scargle1982, Cumming+1999, Zechmeister+2009, VanderPlas+2015, VanderPlas2018, Pearlman+2026a}, the weighted Rayleigh~($Z_{1}^{2}$) periodogram~\citep{Buccheri+1983, Pearlman+2026a}, and the weighted \mbox{$H$-test} periodogram~\citep{deJager+1989, deJager+2010, Pearlman+2026a}. The LS~statistic is most sensitive to approximately sinusoidal modulation. The phase-based $Z_{1}^{2}$~statistic is most sensitive to broad periodic signals that are approximately sinusoidal in phase and dominated by the fundamental harmonic, whereas the \mbox{$H$-test} combines power from multiple harmonics and is therefore better suited to detecting periodic signals with unknown or non-sinusoidal folded profiles~\citep{Pearlman+2026a}. Although the pipeline supports burst-specific weights~\citep{Pearlman+2026a}, we assign uniform weights to all bursts in this work.

We assess the significance of each periodogram peak using source-specific Monte Carlo simulations instead of analytic false-alarm probability~(FAP) calculations. For each source, we generate $N_{\mathrm{sim}}$\,$=$\,$10^{9}$~Monte Carlo realizations under an exposure-modulated homogeneous Poisson null model, in which $N$~independent~ToAs are drawn uniformly in each simulation from time intervals during which the source has non-zero exposure, as determined from the \mbox{CHIME/FRB}~exposure function (see Section~\ref{sec:chime_frb_exposure}). Each null realization preserves the observed burst count and the adopted weighting scheme, while removing any underlying periodic modulation. We compute the~LS, $Z_{1}^{2}$, and \mbox{$H$-test} periodograms for each realization and record the maximum detection statistic across the trial-frequency grid in each periodogram. The resulting~FAPs therefore incorporate the look-elsewhere effect associated with the frequency search. We do not test null models that incorporate burst clustering in this work because of their large parameter space, but potential approaches are discussed in~\citet{Pearlman+2026a}. Since our null model does not incorporate burst clustering, and clustered burst activity can produce excess periodogram power, the resulting significances may decrease if clustering is taken into account.

We apply a multiple-testing correction to obtain the trials-corrected global significance of each candidate (see Section~2.5 in~\citealt{Pearlman+2026a}). The trials factor accounts for the number of detection statistics~($N_{\mathrm{tests}}$), weighting schemes~($N_{\mathrm{weights}}$), and sources~($N_{\mathrm{sources}}$) included in the search. We compute the total number of trials using Equation~(16) in~\citet{Pearlman+2026a}, adopting $N_{\mathrm{tests}}$\,$=$\,3 and $N_{\mathrm{weights}}$\,$=$\,1. We set $N_{\mathrm{sources}}$\,$=$\,1 for the single-source analysis of FRB~20240209A and $N_{\mathrm{sources}}$\,$=$\,19 for the \mbox{CHIME/FRB} Catalog~2 repeater analysis (see Section~\ref{sec:cat2_frb_source_selection}). This yields a total trials factor of $N_{\mathrm{trials}}^{\mathrm{total}}$\,$=$\,3 for the FRB~20240209A analysis and $N_{\mathrm{trials}}^{\mathrm{total}}$\,$=$\,57 for the catalog-wide repeater analysis. Since the three detection statistics are correlated (see Section~2.5 in~\citealt{Pearlman+2026a}), the effective number of independent trials is smaller than these nominal values, and this correction is therefore conservative.

For each source and periodogram, we record the frequency of every peak whose trials-corrected significance exceeds a detection threshold of 3$\sigma$, and we determine the uncertainty on each peak frequency using the statistical method described in Appendix~A of~\citet{Pearlman+2026a}. Since the Poisson null model does not capture all non-periodic structure in the data, formally significant peaks can also arise from sampling artifacts and non-periodic variability. To avoid subjectively selecting which candidates to highlight, we apply the following vetting criteria uniformly to every source:
\begin{enumerate}
	\renewcommand{\labelenumi}{(\roman{enumi})}
	\item \textit{Multi-Statistic Detection:} A repeating~FRB source must have a periodogram peak exceeding the 3$\sigma$ trials-corrected detection threshold in at least two of the three periodograms~(LS, $Z_{1}^{2}$, and \mbox{$H$-test}). The detected power in each periodogram depends on many factors, including the underlying waveform, the sampling of the burst arrival times, and the phase structure of the folded activity profile. However, genuine periodic modulation often produces a signal above this threshold in more than one detection statistic, whereas sampling artifacts and noise fluctuations typically do so in only one, if at all.
	
	\item \textit{Cross-Statistic Period Consistency:} At least two of the above-threshold peaks from different detection statistics must have frequencies that agree within their combined 2$\sigma$ uncertainties~(see Appendix~A of~\citealt{Pearlman+2026a}), either directly or after accounting for a low-order harmonic relation. For example, if one peak lies at~$f_{0}$, another may instead lie at~$k f_{0}$ or~$f_{0}/k$ for an integer~$2$\,$\le$\,$k$\,$\le$\,3. We include harmonic relations because detection statistics that sum power over multiple harmonics, such as the \mbox{$H$-test}, can peak at a subharmonic of the fundamental frequency~(i.e., at an integer multiple of the period) when the number of bursts varies between activity cycles. This requirement rejects cases in which formally significant peaks appear at unrelated frequencies in different detection statistics, since such cases are more likely to arise from noise or sampling artifacts than from a periodic signal.
	
	\item \textit{Activity Phase Concentration:} When folded at the candidate period, the bursts must be concentrated within a well-defined activity phase window rather than being distributed quasi-uniformly across activity phase.
	
\end{enumerate}

We report candidate periodicities only for sources that satisfy all three criteria. For each such source, Table~\ref{table:frb_periodicity_search_summary} lists the candidate activity period measured from the most significant peak satisfying criterion~(ii). When the peaks satisfying criterion~(ii) are harmonically related, we adopt the period at which the bursts are most concentrated in activity phase.

In~\citet{Pearlman+2026a}, we validated this FRB~periodicity search pipeline using simulated data and the $N$\,$=$\,142~ToAs from FRB~20180916B in \mbox{CHIME/FRB} Catalog~2. We recovered FRB~20180916B's well-established \mbox{16.3-day} activity period with high statistical significance in all three periodograms. Table~\ref{table:frb_periodicity_search_summary} also includes the refined activity period and activity period derivative of FRB~20180916B reported in~\citet{Pearlman+2026a}.


\subsection{\mbox{CHIME/FRB} Catalog~2: Repeating~FRB \\ Source Selection}
\label{sec:cat2_frb_source_selection}

For the catalog-wide search, we do not analyze the burst~ToAs of every repeating~FRB source in \mbox{CHIME/FRB} Catalog~2. Many repeaters have too few detected bursts, or burst arrival times that span too small a fraction of the observing baseline, to yield informative periodicity constraints. We therefore introduce the following heuristic criterion to identify the subset of sources for which a periodicity search is most likely to be informative.

The duty cycle~($\delta$) of a periodic repeating~FRB source is defined as
\begin{equation}
	\delta = \frac{W_{\mathrm{eff}}}{P_{\mathrm{activity}}},
	\label{eqn:duty_cycle}
\end{equation}
where 0\,$\le$\,$\delta$\,$\le$\,1, $W_{\mathrm{eff}}$ is the effective width of the activity window (i.e.,~the time interval during which the source is active and detectable), and $P_{\mathrm{activity}}$ is the putative activity period. If the bursts occur randomly in time, the probability that a single~ToA falls within the active phase window is
\begin{equation}
	P_{1} = \delta.
	\label{eqn:prob_1_burst}
\end{equation}
Assuming the burst arrival times are independent, the probability that all $N$~ToAs fall within the active phase window by chance is
\begin{equation}
	P(N) = \prod_{i=1}^{N} P_{i} = \delta^{N},
	\label{eqn:prob_n_bursts}
\end{equation}
where $P_{i}$\,$=$\,$\delta$ for 1\,$\le$\,$i$\,$\le$\,$N$. 
Requiring $P(N)$ to be smaller than a chosen~FAP, $P_{\mathrm{FAP}}$, yields an approximate lower limit on the number of~ToAs needed to detect a periodic signal:
\begin{equation}
	 N \geq \left\lfloor \frac{\log{P_{\mathrm{FAP}}}}{\log{\delta}} \right\rfloor,
	\label{eqn:num_toas_criterion}
\end{equation}
where $\lfloor \cdot \rfloor$ denotes the floor function. We round down, rather than up, in Equation~(\ref{eqn:num_toas_criterion}) to avoid excluding potentially informative sources near the chosen detection threshold.

We adopt a nominal duty cycle of $\delta$\,$=$\,0.6 in Equation~(\ref{eqn:num_toas_criterion}), motivated by timing observations of FRB~20180916B~\citep{CHIME+2020b, Pastor-Marazuela+2021, Pleunis+2021b, Sand+2023, Pearlman+2026a} and FRB~20121102A~\citep{Rajwade+2020b, Cruces+2021, Braga+2025, Gouiffes+2025, Wang+2025}. The \mbox{16.3-day} periodic activity of FRB~20180916B has a duty cycle of $\delta$\,$\approx$\,0.3 in the \mbox{400--800}\,MHz band~\citep{Sand+2023}, while FRB~20121102A shows evidence of a possible \mbox{160-day} activity period with a duty cycle of $\delta$\,$\approx$\,0.6 at 1.4\,GHz~\citep{Rajwade+2020b}. Since bursts from a source with a larger duty cycle are more likely to fall within the activity window by chance, more~ToAs are needed to reach a given~FAP. We therefore adopt the larger of these two values, $\delta$\,$=$\,0.6, so that the minimum number of burst~ToAs required for inclusion in our sample is not underestimated for sources with broad activity windows.

We require that candidate periodic signals exceed a trials-corrected statistical significance of~3$\sigma$, corresponding to a one-sided~FAP of
\begin{equation}
	P_{3\sigma} = \frac{1}{2}\left[1-\mathrm{erf}\!\left( \frac{3}{\sqrt{2}} \right)\right] \approx 1.35 \times 10^{-3}.
	\label{eqn:prob_3sigma}
\end{equation}
Substituting $P_{\mathrm{FAP}}$\,$=$\,$P_{3\sigma}$ and $\delta$\,$=$\,0.6 into Equation~(\ref{eqn:num_toas_criterion}) yields $N$\,$\geq$\,12. We therefore analyze all repeating~FRB sources in \mbox{CHIME/FRB} Catalog~2 that satisfy this criterion, with the exception of FRB~20210601A. Although FRB~20210601A has 13~cataloged~ToAs, they span only $\sim$2\,minutes (see Figure~\ref{fig:frb_toa_timeline} and~\citealt{Mckinven+2026}), which makes a search for periodicities longer than $P_{\mathrm{min}}$\,$=$\,10\,days infeasible (see Section~\ref{sec:frb_periodicity_search_pipeline_overview}). This yields a final sample of $N_{\mathrm{sources}}$\,$=$\,19~sources (see Table~\ref{table:frb_periodicity_search_summary}). We search the cataloged~ToAs of each source individually for periodicity, using its source-specific \mbox{CHIME/FRB} exposure function, but treat all 19~sources as a single catalog-wide analysis when computing the trials correction (see Section~\ref{sec:frb_periodicity_search_pipeline_overview}).

We emphasize that $\delta$\,$=$\,0.6 is a heuristic choice, and individual sources may exhibit larger duty cycles. For example, we measure a duty cycle of $\delta$\,$=$\,0.89 for FRB~20240209A (see Section~\ref{sec:frb20240209a_results}), which would require $N$\,$\geq$\,56~ToAs to reach an~FAP below~$P_{3\sigma}$. Our sample of 155~bursts from FRB~20240209A comfortably satisfies this criterion.


\section{Results}
\label{sec:results}

In this section, we present the results of our FRB~periodicity search. Unless otherwise stated, all reported uncertainties correspond to 1$\sigma$ confidence intervals. All quoted detection significances of candidate periodic signals are determined from Monte Carlo simulations, corrected for the number of trials, and expressed as one-sided Gaussian-equivalent values (see Section~\ref{sec:frb_periodicity_search_pipeline_overview} and~\citealt{Pearlman+2026a}). They represent the global significance of each candidate relative to the null hypothesis that the burst arrival times are drawn from an exposure-modulated homogeneous Poisson process (i.e.,~a process with neither an underlying periodicity nor temporal clustering), conditioned on the observed number of bursts.

We interpret the candidate periodicities identified in this work as arising from the cyclic bursting activity of each source, rather than from coherent pulsed emission of the kind observed from rotating radio pulsars and magnetars~(e.g.,~\citealt{Manchester+2005, McLaughlin+2006, Pearlman+2018a}). In addition to recovering the known \mbox{16.3-day} activity period of FRB~20180916B~\citep{Pearlman+2026a}, we identify candidate periodicities in~4 of the 20~sources searched. We note that the absence of a significant periodic signal from the remaining sources does not rule out underlying periodic behavior, which may be revealed by future observations. Our search results are summarized in Table~\ref{table:frb_periodicity_search_summary}.


\subsection{FRB~20240209A}
\label{sec:frb20240209a_results}

\begin{figure*}
	\centering
	\includegraphics[scale=0.83]{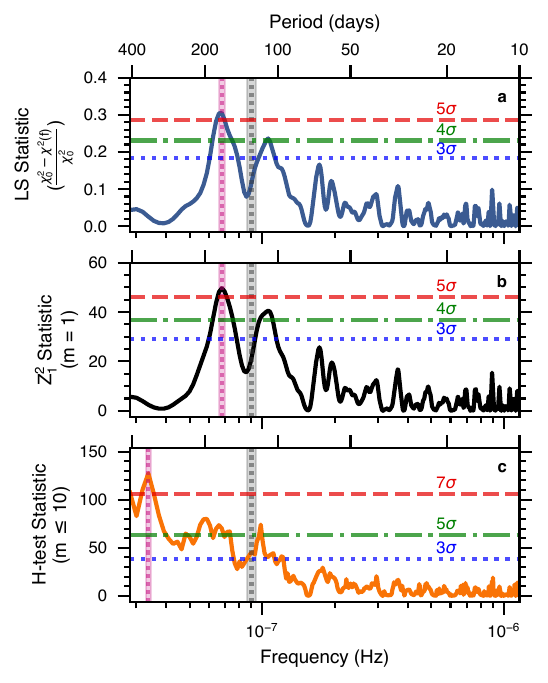} \hfill
	\includegraphics[scale=0.83]{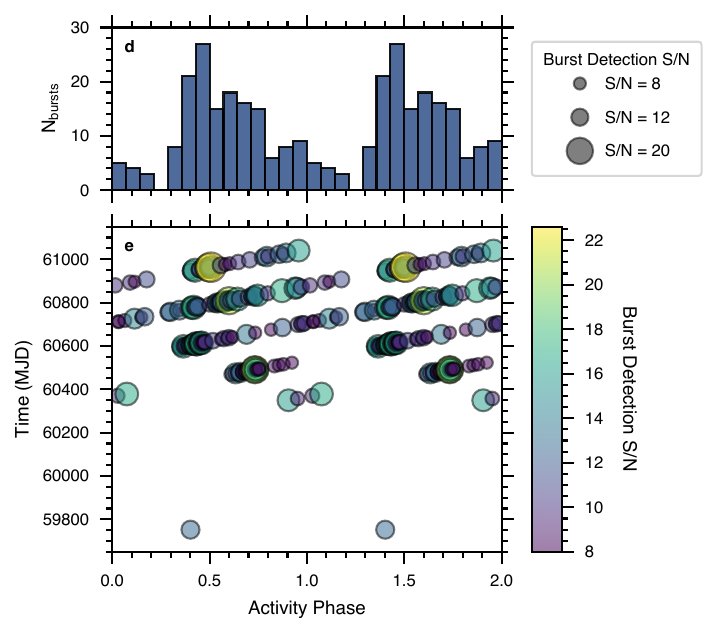}
	\caption{Periodograms and folded activity profile of FRB~20240209A. Panels~a, b, and~c show the~LS, $Z_{1}^{2}$, and \mbox{$H$-test} periodograms, respectively, computed from the $N$\,$=$\,155~burst arrival times listed in Table~\ref{table:frb20240209a_burst_log}. In each panel, the magenta vertical dotted line marks the frequency of the highest peak, and the magenta shaded region denotes its 1$\sigma$ uncertainty. The horizontal blue dotted, green dot-dashed, and red dashed lines indicate trials-corrected detection thresholds derived from Monte Carlo simulations. The gray vertical dotted line and shaded region mark the \mbox{128.5-day} activity period reported by~\citet{Pal2025} and its uncertainty. Panel~d shows the activity profile obtained by folding the burst arrival times modulo the \mbox{170.4-day} candidate activity period identified in this work, using a reference epoch of $t_{\mathrm{ref}}$\,$=$\,MJD\,60706.27987. Panel~e shows the distribution of the bursts as a function of time and activity phase. Each circle represents one burst, with its color and size scaled by the detection~\mbox{S/N}.}
	\label{fig:frb20240209a_periodogram_activity_profile}
\end{figure*}

We searched for periodicities from FRB~20240209A using the $N$\,$=$\,155 burst arrival times listed in Table~\ref{table:frb20240209a_burst_log}. The resulting~LS, $Z_{1}^{2}$, and \mbox{$H$-test} periodograms are shown in Figure~\ref{fig:frb20240209a_periodogram_activity_profile}\mbox{a--c}, together with trials-corrected detection thresholds derived from Monte Carlo simulations. The~LS and $Z_{1}^{2}$~periodograms yield consistent candidate activity periods of 170.4\,$\pm$\,4.7\,days and 170.4\,$\pm$\,4.8\,days, respectively, each with a trials-corrected statistical significance of 5.3$\sigma$. The strongest peak in the \mbox{$H$-test} periodogram occurs at 342.8\,$\pm$\,8.3\,days (7.8$\sigma$), consistent with the second subharmonic~($f_{0}/2$) of the fundamental frequency, $f_{0}$\,$=$\,1/(170.4\,days).

Figure~\ref{fig:frb20240209a_periodogram_activity_profile}d shows the activity profile obtained by folding the burst arrival times modulo the \mbox{170.4-day} candidate activity period, and Figure~\ref{fig:frb20240209a_periodogram_activity_profile}e shows the distribution of the bursts in time and activity phase. The burst activity phases yield a duty cycle of $\delta$\,$=$\,0.89\,$\pm$\,0.02, with enhanced activity between phases $\phi$\,$=$\,\mbox{0.34--0.76}. Although bursts are detected across most of the activity cycle, 75\% of them occur within this enhanced activity window, which spans 42\% of the activity cycle.

\begin{figure}
	\centering
	\includegraphics[width=\linewidth, trim=6bp 6bp 6bp 6bp, clip]{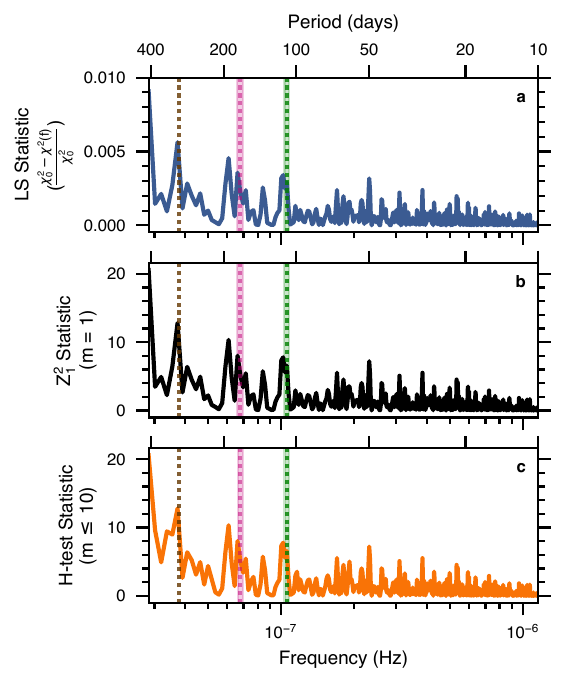}
	\caption{Periodograms of the exposure function of FRB~20240209A. Panels~a, b, and~c show the~LS, $Z_{1}^{2}$, and \mbox{$H$-test} periodograms, respectively, computed from the exposure data shown in Figure~\ref{fig:frb_source_exposure}d. In each panel, the magenta vertical dotted line and shaded region mark the frequency and 1$\sigma$ uncertainty of the \mbox{170.4-day} candidate activity period identified for FRB~20240209A in this work. The green vertical dotted line and shaded region similarly mark the aliased, secondary \mbox{109.5-day} peak and its 1$\sigma$ uncertainty. The brown vertical dotted line marks the inferred window-function frequency, corresponding to a period of 306.6\,days, which gives rise to the secondary peak at 109.5\,days. Significant power is present at the window-function frequency in all three periodograms of FRB~20240209A's exposure function.}
	\label{fig:frb20240209a_periodogram_exposure}
\end{figure}

The~LS and $Z_{1}^{2}$~periodograms also show a statistically significant secondary
peak at 109.5\,$\pm$\,2.4\,days and 109.5\,$\pm$\,2.7\,days, with significances of
4.1$\sigma$ and 4.4$\sigma$, respectively. To test whether this peak is an alias introduced by the observational window function, we computed the~LS, $Z_{1}^{2}$, and \mbox{$H$-test} periodograms of FRB~20240209A's exposure function over the full observing campaign (see Figure~\ref{fig:frb20240209a_periodogram_exposure}). Since the exposure pattern remained broadly stable throughout the campaign (see Figure~\ref{fig:frb_source_exposure}d), these periodograms also characterize the window function over the interval spanned by the detected bursts. The periodograms of the exposure reveal significant power at a window-function period of 306.6\,days, which is unrelated to the burst activity itself.

A peak near 306.6\,days also appears in the exposure periodograms of other repeaters spanning a wide range of declinations and right ascensions, indicating that this feature is not source-specific. Although we do not identify a single definitive cause, the peak power is reduced when outliers are removed from the exposure function (e.g.,~by excluding times at which the exposure deviates by more than 3$\sigma$ from its mean). This suggests that the feature arises, at least in part, from a small number of atypical exposures that introduce long-timescale structure into the exposure function.

Since the observed Fourier spectrum is effectively the convolution of the intrinsic signal spectrum with the spectral window (the Fourier transform of the window function describing the observing cadence), a peak in the spectral window at frequency~$f_{\mathrm{w}}$ can produce aliases of the fundamental signal at frequencies~$f_{\mathrm{alias}}$\,$=$\,$|f_{0}$\,$\pm$\,$f_{\mathrm{w}}|$. Taking~$f_{\mathrm{w}}$\,$=$\,1/(306.6\,days), the alias at~$f_{0}$\,$+$\,$f_{\mathrm{w}}$ corresponds to a period of~109.5\,days, consistent with the observed secondary peak. The \mbox{109.5-day} peak is therefore likely an exposure-induced alias rather than an independent periodicity.

No corresponding alias is observed at~$f_{0}$\,$-$\,$f_{\mathrm{w}}$ (i.e.,~at a period of~383.6\,days). This asymmetry can arise because the power at each alias frequency is not determined solely by the spectral-window peak at~$f_{\mathrm{w}}$. Each alias also receives contributions from the negative-frequency counterpart of the signal, which is present for any real-valued signal, and from harmonics arising from the non-sinusoidal shape of the periodic activity. These contributions differ in amplitude and phase at the two alias frequencies, so they can enhance the alias power at one frequency while suppressing it at the other~\citep{Dawson+2010}.

The exposure periodograms also show residual power above the mean near~$f_{0}$. However, our Monte Carlo simulations draw arrival times from the source-specific exposure function, so the trials-corrected detection thresholds shown in Figure~\ref{fig:frb20240209a_periodogram_activity_profile}\mbox{a--c} account for the structure in the exposure spectrum at every frequency searched. The 5.3$\sigma$ significance of the \mbox{170.4-day} candidate activity period is therefore robust to aliasing structure imposed by the window function.

We find no statistically significant evidence for the activity period of 128.5\,$\pm$\,6\,days reported by~\citet{Pal2025}. At this period (marked by the gray vertical dotted line and shaded region in Figure~\ref{fig:frb20240209a_periodogram_activity_profile}\mbox{a--c}), the~LS and $Z_{1}^{2}$ periodograms yield significances of only 0.7$\sigma$ and 1.6$\sigma$, respectively, both well below the 3$\sigma$ trials-corrected detection threshold. Moreover, our \mbox{170.4-day} candidate activity period is inconsistent with the \mbox{128.5-day} period at the 5.5$\sigma$~level, based on the difference between our LS~measurement and the most precise value reported in Table~1 of~\citet{Pal2025}. Our data therefore do not support a constant \mbox{128.5-day} activity period for FRB~20240209A.

\begin{figure}
	\centering
	\includegraphics[width=\linewidth, trim=6bp 6bp 6bp 6bp, clip]{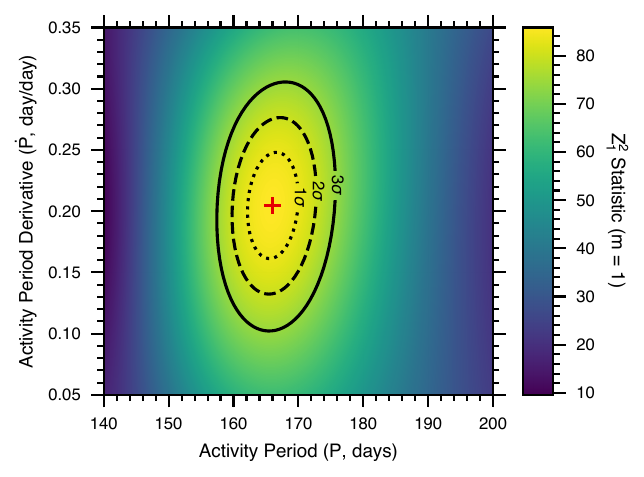}
	\caption{Joint measurements of the activity period~($P$) and activity period derivative~($\dot{P}$) of FRB~20240209A, obtained by maximizing the $Z_{1}^{2}$~statistic over a grid of \mbox{($P$, $\dot{P}$)}~values using the method described in Appendix~B of~\citet{Pearlman+2026a}. The color scale shows the $Z_{1}^{2}$~statistic for each \mbox{($P$, $\dot{P}$)} pair, evaluated at a reference epoch of $t_{\mathrm{ref}}$\,$=$\,MJD\,60712.47454. The red plus symbol marks the \mbox{best-fit} values. The black contours correspond to lines of constant $\Delta Z_{1}^{2}$\,$=$\,$Z_{1,\mathrm{max}}^{2} - Z_{1}^{2}$. The black dotted, dashed, and solid lines enclose the~1$\sigma$, 2$\sigma$, and 3$\sigma$~joint confidence regions, corresponding to $\Delta Z_{1}^{2}$\,$=$\,2.30, 6.18, and 11.83, respectively, for two parameters of interest under the likelihood-ratio approximation (Wilks' theorem;~\citealt{Wilks1938, Cowan+2011}).}
	\label{fig:frb20240209a_ppdot}
\end{figure}

Next, we searched for possible temporal evolution of the candidate activity period by computing the $Z_{1}^{2}$~statistic over a two-dimensional grid of activity periods~($P$) and activity period derivatives~($\dot{P}$), following the method described in Appendix~B of~\citet{Pearlman+2026a}. The $Z_{1}^{2}$~statistic is well suited to event-arrival-time data with a broad folded profile and allows confidence regions to be derived directly from~$\Delta Z_{1}^{2}$. We excluded the first burst (B1~in Table~\ref{table:frb20240209a_burst_log}) from the \mbox{$(P, \dot{P})$}~analysis, since the large gap between the first and second~ToAs introduces a cycle-count ambiguity that produces multiple, nearly equally significant maxima in the \mbox{$(P, \dot{P})$}~plane. This ambiguity outweighs the benefit of the longer temporal baseline. The resulting $Z_{1}^{2}$~surface is shown in Figure~\ref{fig:frb20240209a_ppdot}, where the dotted, dashed, and solid contours denote the 1$\sigma$, 2$\sigma$, and 3$\sigma$ two-parameter confidence regions, respectively.

The joint confidence contours in the \mbox{$(P, \dot{P})$}~plane are nearly aligned with the coordinate axes, indicating weak covariance between~$P$ and~$\dot{P}$. We quantify this covariance non-parametrically from the weighted second central moments of the $Z_{1}^{2}$~surface, adopting the relative likelihood $\mathcal{L}(P, \dot{P})$\,$\propto$\,$\exp[-\Delta Z_{1}^{2}(P, \dot{P})/2]$ as the weight over the $(P, \dot{P})$~grid, where $\Delta Z_{1}^{2}$ is measured relative to the maximum. This yields a correlation coefficient of $\rho$\,$=$\,$\mathrm{cov}(P, \dot{P})/(\sigma_{P}\,\sigma_{\dot{P}})$\,$=$\,0.13, where $\mathrm{cov}(P, \dot{P})$ is the weighted second central cross-moment between~$P$ and~$\dot{P}$, and $\sigma_{P}$ and $\sigma_{\dot{P}}$ are the square roots of the corresponding weighted variances.

\begin{figure}
	\centering
	\includegraphics[width=\linewidth, trim=6bp 6bp 6bp 6bp, clip]{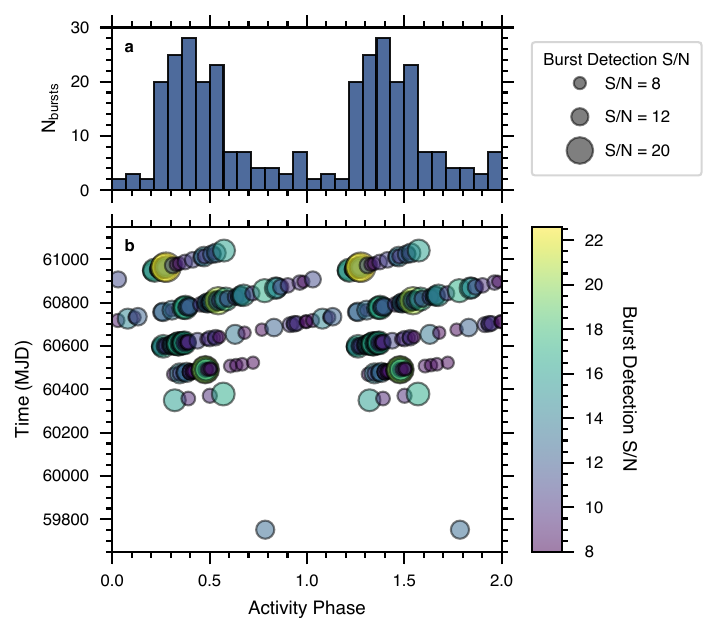}
	\caption{Folded activity profile of FRB~20240209A, computed from the same bursts as in Figure~\ref{fig:frb20240209a_periodogram_activity_profile}. The activity profile is obtained using a reference epoch of $t_{\mathrm{ref}}$\,$=$\,MJD\,60712.47454 and the candidate activity period and activity period derivative pair~\mbox{($P$, $\dot{P}$)\,$=$\,(166.0\,days, 0.20\,day\,day$^{\mathrm{-1}}$)} that maximizes the $Z_{1}^{2}$~statistic in Figure~\ref{fig:frb20240209a_ppdot}. Panel~a shows the activity profile. Panel~b shows the distribution of the bursts as a function of time and activity phase. Each circle represents one burst, with its color and size scaled by the detection~\mbox{S/N}.}
	\label{fig:frb20240209a_activity_profile_ppdot}
\end{figure}

From the location of the maximum $Z_{1}^{2}$~statistic on a fine $(P, \dot{P})$~grid, we obtain $P$\,$=$\,166.0$^{\mathrm{+4.0}}_{\mathrm{-3.8}}$\,days and $\dot{P}$\,$=$\,0.20$^{\mathrm{+0.04}}_{\mathrm{-0.04}}$\,day\,day$^{\mathrm{-1}}$, with a reference epoch of $t_{\mathrm{ref}}$\,$=$\,MJD\,60712.47454. The uncertainties are determined by projecting the 1$\sigma$ two-parameter confidence region onto each coordinate axis (see Appendix~B of~\citealt{Pearlman+2026a}). This period agrees within~1$\sigma$ with the \mbox{170.4-day} candidate activity period identified in the~LS and $Z_{1}^{2}$~periodograms, indicating that the candidate~$\dot{P}$ is associated with the same periodicity. Figure~\ref{fig:frb20240209a_activity_profile_ppdot} shows the activity profile obtained by folding the burst arrival times with these values of $P$ and $\dot{P}$, which exhibits enhanced activity between phases $\phi$\,$=$\,\mbox{0.20--0.59}. This window contains 78\% of the bursts while spanning 39\% of the activity cycle. Compared with the constant-period activity profile in Figure~\ref{fig:frb20240209a_periodogram_activity_profile}\mbox{d--e}, incorporating this non-zero~$\dot{P}$ concentrates the bulk of the bursts more tightly in activity phase, while redistributing a small number of bursts into previously unoccupied phases.

We regard this candidate activity period derivative as very tentative and do not claim a detection. The quoted uncertainties describe the location and shape of the $Z_{1}^{2}$~maximum under the assumption that a~$\dot{P}$ term belongs in the timing model. They do not indicate whether a non-zero~$\dot{P}$ is required by the data. In addition, the burst arrival times used in the \mbox{$(P, \dot{P})$}~analysis span only $\sim$4.2~activity cycles, over which the best-fit~$\dot{P}$ corresponds to a cumulative phase drift of $\sim$0.4~cycles relative to a constant period, which is comparable to the width of the enhanced activity window. Such a drift could plausibly arise from burst clustering or other cycle-to-cycle variations in the activity of the source, rather than from genuine evolution of an activity period. Since these effects are poorly constrained for FRB~20240209A, any~FAP derived from simulations would depend strongly on the assumed clustering and activity model. Continued monitoring of this source will provide a more direct test, since the constant-period and non-zero-$\dot{P}$ timing models predict increasingly divergent times for future windows of enhanced activity.


\subsection{FRB~20190208A}
\label{sec:frb20190208a_results}

\begin{figure*}
	\centering
	\includegraphics[scale=0.83]{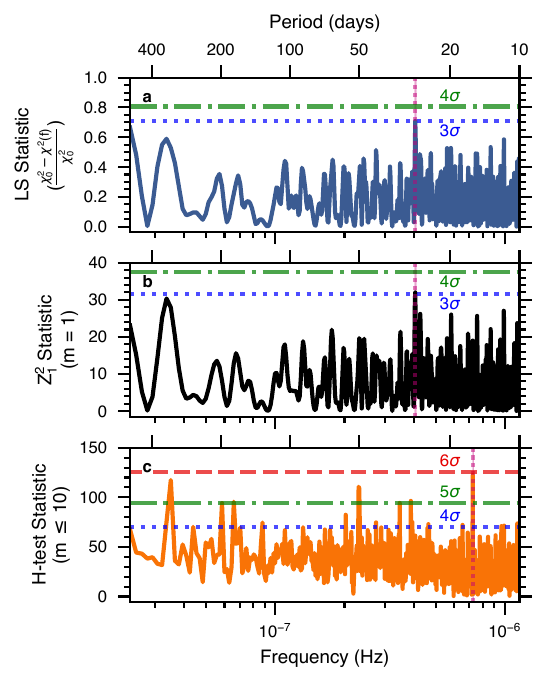} \hfill
	\includegraphics[scale=0.83]{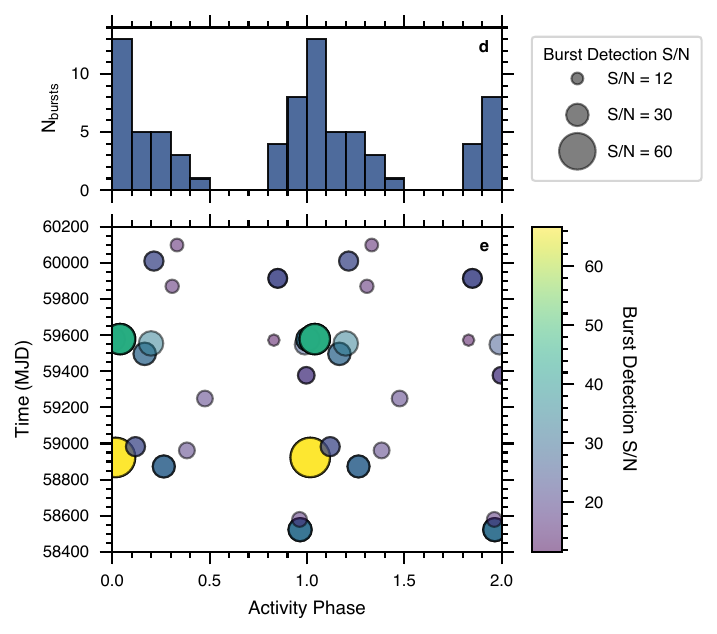}
	\caption{Periodograms and folded activity profile of FRB~20190208A. Panels~a, b, and~c show the~LS, $Z_{1}^{2}$, and \mbox{$H$-test} periodograms, respectively, computed from the $N$\,$=$\,39~burst arrival times in \mbox{CHIME/FRB} Catalog~2. In each panel, the magenta vertical dotted line marks the frequency of the highest peak, and the magenta shaded region denotes its 1$\sigma$ uncertainty. The horizontal blue dotted, green dot-dashed, and red dashed lines indicate trials-corrected detection thresholds derived from Monte Carlo simulations. Panel~d shows the activity profile obtained by folding the burst arrival times modulo the \mbox{28.46-day} candidate activity period identified in this work, using a reference epoch of $t_{\mathrm{ref}}$\,$=$\,MJD\,59320.57734. Panel~e shows the distribution of the bursts as a function of time and activity phase. Each circle represents one burst, with its color and size scaled by the detection~\mbox{S/N}.}
	\label{fig:frb20190208a_periodogram_activity_profile}
\end{figure*}

Our periodicity search for FRB~20190208A uses the arrival times of the $N$\,$=$\,39~bursts in \mbox{CHIME/FRB}~Catalog~2. Figure~\ref{fig:frb20190208a_periodogram_activity_profile}\mbox{a--c} shows the resulting~LS, $Z_{1}^{2}$, and \mbox{$H$-test} periodograms, along with trials-corrected detection thresholds obtained from Monte Carlo simulations. The $Z_{1}^{2}$~periodogram shows a peak at 28.46\,$\pm$\,0.08\,days, with a statistical significance of 3.1$\sigma$. The LS~periodogram recovers a consistent period of 28.46\,$\pm$\,0.07\,days at 3.0$\sigma$.

The strongest peak in the \mbox{$H$-test} periodogram occurs at 15.99\,$\pm$\,0.01\,days, with a significance of 5.9$\sigma$. This period is not harmonically related to the \mbox{28.46-day} period and is not recovered by the~LS or $Z_{1}^{2}$~periodograms, so it does not satisfy criterion~(ii) in Section~\ref{sec:frb_periodicity_search_pipeline_overview}. We also note that the 16\textsuperscript{th}~harmonic of the \mbox{15.99-day} period falls near one cycle per day, the frequency at which \mbox{CHIME/FRB} observes the source. We therefore adopt 28.46\,days, which is recovered consistently by both the~LS and $Z_{1}^{2}$~statistics, as the candidate activity period for FRB~20190208A.

Folding the burst arrival times modulo the \mbox{28.46-day} candidate activity period yields the activity profile shown in Figure~\ref{fig:frb20190208a_periodogram_activity_profile}d. Figure~\ref{fig:frb20190208a_periodogram_activity_profile}e shows the distribution of the bursts in time and activity phase. We measure a duty cycle of $\delta$\,$=$\,0.646\,$\pm$\,0.002 from the folded activity profile in Figure~\ref{fig:frb20190208a_periodogram_activity_profile}.


\subsection{FRB~20200926A}
\label{sec:frb20200926a_results}

\begin{figure*}
	\centering
	\includegraphics[scale=0.83]{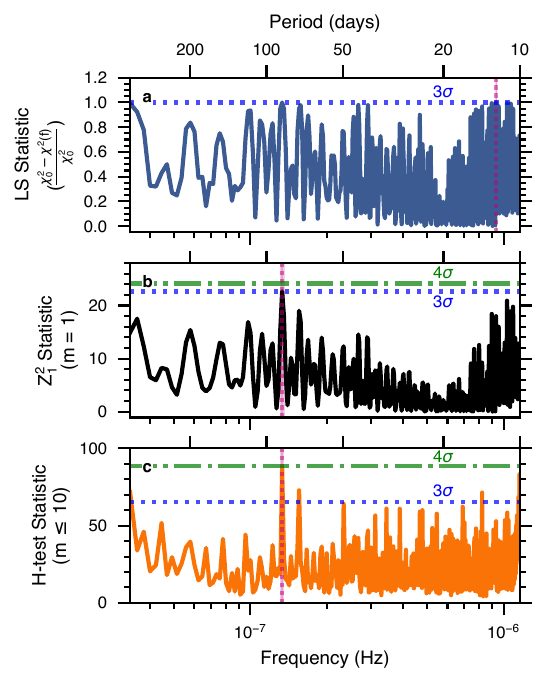} \hfill
	\includegraphics[scale=0.83]{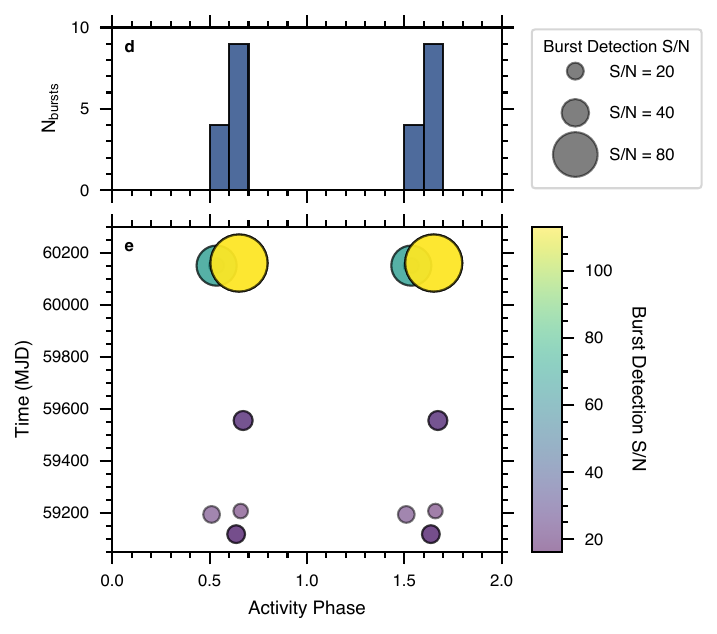}
	\caption{Periodograms and folded activity profile of FRB~20200926A. Panels~a, b, and~c show the~LS, $Z_{1}^{2}$, and \mbox{$H$-test} periodograms, respectively, computed from the $N$\,$=$\,13~burst arrival times in \mbox{CHIME/FRB} Catalog~2. In each panel, the magenta vertical dotted line marks the frequency of the highest peak, and the magenta shaded region denotes its 1$\sigma$ uncertainty. The horizontal blue dotted and green dot-dashed lines indicate trials-corrected detection thresholds derived from Monte Carlo simulations. Panel~d shows the activity profile obtained by folding the burst arrival times modulo the \mbox{86.7-day} candidate activity period identified in this work, using a reference epoch of $t_{\mathrm{ref}}$\,$=$\,MJD\,59756.81809. Panel~e shows the distribution of the bursts as a function of time and activity phase. Each circle represents one burst, with its color and size scaled by the detection~\mbox{S/N}.}
	\label{fig:frb20200926a_periodogram_activity_profile}
\end{figure*}

FRB~20200926A has only $N$\,$=$\,13~bursts in \mbox{CHIME/FRB}~Catalog~2, the smallest sample among the four sources with candidate periodicities. The~LS, $Z_{1}^{2}$, and \mbox{$H$-test} periodograms computed from these burst arrival times are displayed in Figure~\ref{fig:frb20200926a_periodogram_activity_profile}\mbox{a--c}, together with trials-corrected detection thresholds derived from Monte Carlo simulations. The most significant peak appears in the \mbox{$H$-test} periodogram at 86.7\,$\pm$\,0.5\,days, with a statistical significance of 4.0$\sigma$. The $Z_{1}^{2}$~periodogram recovers a consistent period of 86.7\,$\pm$\,1.1\,days at 3.0$\sigma$.

The highest peak in the LS~periodogram is located at 12.40\,$\pm$\,0.02\,days, with a significance of 2.9$\sigma$, just below our 3$\sigma$ detection threshold. This period is consistent with the seventh harmonic~(7$f_{0}$) of the fundamental frequency, $f_{0}$\,$=$\,1/(86.7\,days). The \mbox{12.4-day} peak may reflect power associated with the \mbox{86.7-day} period that is spread into higher harmonics by the narrow activity profile, rather than an independent periodicity. Supporting this interpretation, the LS~periodogram also shows a slightly lower peak at 86.7\,$\pm$\,0.9\,days, likewise at 2.9$\sigma$. We therefore adopt the \mbox{86.7-day} period, which is the dominant peak in both the \mbox{$H$-test} and $Z_{1}^{2}$~periodograms, as the candidate activity period for FRB~20200926A.

When the burst arrival times are folded modulo the \mbox{86.7-day} candidate activity period, the bursts are confined to a narrow range of activity phases. Figure~\ref{fig:frb20200926a_periodogram_activity_profile}e shows the distribution of the bursts in time and activity phase. From the folded activity profile in Figure~\ref{fig:frb20200926a_periodogram_activity_profile}, we measure a duty cycle of $\delta$\,$=$\,0.162\,$\pm$\,0.001.


\subsection{FRB~20220529A}
\label{sec:frb20220529a_results}

\begin{figure*}
	\centering
	\includegraphics[scale=0.83]{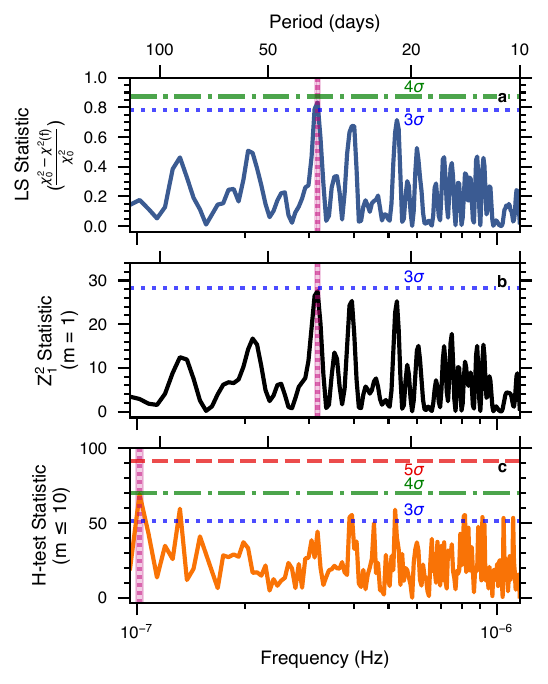} \hfill
	\includegraphics[scale=0.83, trim=0bp 0.42bp 0bp 0bp]{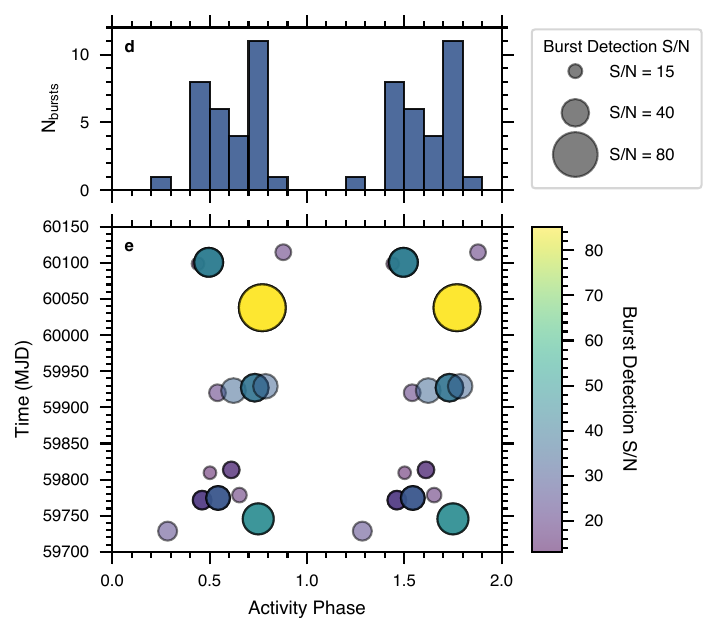}
	\caption{Periodograms and folded activity profile of FRB~20220529A. Panels~a, b, and~c show the~LS, $Z_{1}^{2}$, and \mbox{$H$-test} periodograms, respectively, computed from the $N$\,$=$\,31~burst arrival times in \mbox{CHIME/FRB} Catalog~2. In each panel, the magenta vertical dotted line marks the frequency of the highest peak, and the magenta shaded region denotes its 1$\sigma$ uncertainty. The horizontal blue dotted, green dot-dashed, and red dashed lines indicate trials-corrected detection thresholds derived from Monte Carlo simulations. Panel~d shows the activity profile obtained by folding the burst arrival times modulo the \mbox{36.4-day} candidate activity period identified in this work, using a reference epoch of $t_{\mathrm{ref}}$\,$=$\,MJD\,59900.48736. Panel~e shows the distribution of the bursts as a function of time and activity phase. Each circle represents one burst, with its color and size scaled by the detection~\mbox{S/N}.}
	\label{fig:frb20220529a_periodogram_activity_profile}
\end{figure*}

\mbox{CHIME/FRB}~Catalog~2 contains $N$\,$=$\,31~bursts from FRB~20220529A. Figure~\ref{fig:frb20220529a_periodogram_activity_profile}\mbox{a--c} shows the~LS, $Z_{1}^{2}$, and \mbox{$H$-test} periodograms computed from these burst arrival times, along with trials-corrected detection thresholds obtained from Monte Carlo simulations. The strongest peak in the LS~periodogram occurs at 36.4\,$\pm$\,0.5\,days, with a statistical significance of 3.5$\sigma$. The $Z_{1}^{2}$~periodogram recovers the same period (36.4\,$\pm$\,0.5\,days) at a significance of 2.8$\sigma$, slightly below our 3$\sigma$ threshold.

In the \mbox{$H$-test} periodogram, the dominant peak is instead located at 113.7\,$\pm$\,2.7\,days, with a significance of 4.0$\sigma$. This period is 3.12\,$\pm$\,0.08 times the \mbox{36.4-day} period, consistent at the 1.4$\sigma$ level with the third subharmonic~($f_{0}/3$) of the fundamental frequency, $f_{0}$\,$=$\,1/(36.4\,days). When folded modulo 113.7\,days, the bursts fall in separate groups spaced by approximately one-third of a cycle, as expected for this subharmonic. We therefore adopt 36.4\,days as the candidate activity period for FRB~20220529A.

Folding the burst arrival times modulo the \mbox{36.4-day} candidate activity period yields the activity profile shown in Figure~\ref{fig:frb20220529a_periodogram_activity_profile}d. The distribution of the bursts in time and activity phase is shown in Figure~\ref{fig:frb20220529a_periodogram_activity_profile}e. The bursts occupy a broad activity window, with a duty cycle of $\delta$\,$=$\,0.594\,$\pm$\,0.008.

\begin{figure}
	\centering
	\includegraphics[width=\linewidth, trim=6bp 6bp 6bp 6bp, clip]{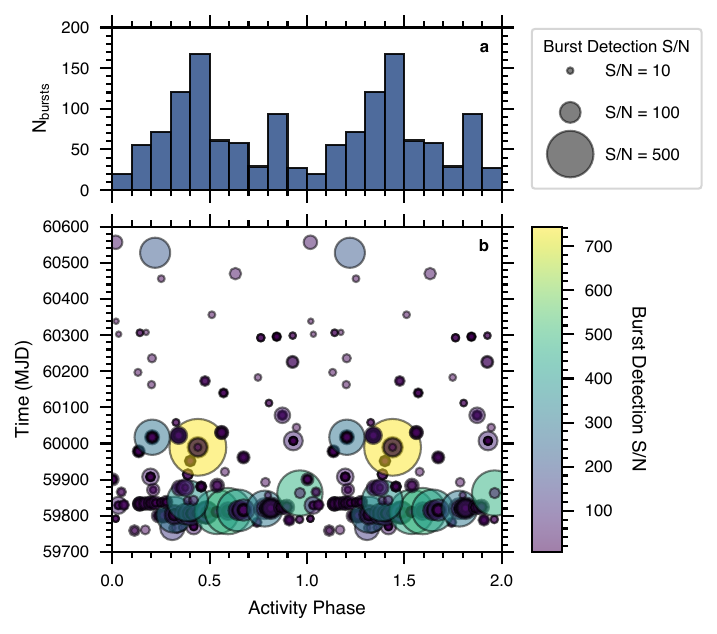}
	\caption{Folded activity profile of FRB~20220529A from the $N$\,$=$\,704~burst arrival times measured with FAST and reported in~\citet{Li+2026a}. Panel~a shows the activity profile obtained by folding the burst arrival times modulo the \mbox{36.4-day} candidate activity period identified in this work, using the same reference epoch~($t_{\mathrm{ref}}$\,$=$\,MJD\,59900.48736) as in Figure~\ref{fig:frb20220529a_periodogram_activity_profile}\mbox{d--e}. Panel~b shows the distribution of the bursts as a function of time and activity phase. Each circle represents one burst, with its color and size scaled by the detection~\mbox{S/N}.}
	\label{fig:frb20220529a_activity_profile_fast}
\end{figure}

Observations with the Five-hundred-meter Aperture Spherical radio Telescope~(FAST) provide an independent test of this candidate activity period. Since FAST and \mbox{CHIME/FRB} observe the source at different times, these observations can also help assess whether the period could instead be an alias introduced by the \mbox{CHIME/FRB} window function~(see Section~\ref{sec:activity_period_aliasing}). \citet{Li+2026a} reported~ToAs for 704~bursts detected with~FAST between 1.0 and 1.5\,GHz. These bursts span 2.2\,years, or 21.9\,cycles of the \mbox{36.4-day} candidate activity period, but 92\% of them were detected within the first 263.3\,days of the campaign (i.e.,~within only 7.2~putative activity cycles). Applying our FRB~periodicity search pipeline (see Section~\ref{sec:frb_periodicity_search_pipeline_overview} and~\citealt{Pearlman+2026a}) to these FAST~ToAs, we find no significant evidence of the \mbox{36.4-day} period. However, when the FAST~ToAs are folded at this period using the same reference epoch as in Figure~\ref{fig:frb20220529a_periodogram_activity_profile}\mbox{d--e}, approximately 76\% of the bursts lie between activity phases~0.1 and~0.7 (see Figure~\ref{fig:frb20220529a_activity_profile_fast}).

Two factors, which need not be mutually exclusive, may explain why our FRB~periodicity search of the FAST~data did not recover the \mbox{36.4-day} candidate activity period. First, since most of the FAST~bursts were detected within the first third of the campaign, the \mbox{36.4-day} period is well sampled over only a few consecutive cycles, which limits the statistical power of the search even though the campaign formally spans many cycles. The irregular observing cadence of~FAST can also produce uneven coverage in activity phase, which may distort or suppress a periodic signal. The folded profile in Figure~\ref{fig:frb20220529a_activity_profile_fast} may therefore partly reflect the phases at which the source was observed rather than intrinsic variations in burst activity. Second, the periodic activity could be frequency-dependent (e.g.,~through mechanisms analogous to pulsar radius-to-frequency mapping;~\citealt{Komesaroff1970, Cordes1978, Beniamini+2020a}), in which case the \mbox{CHIME/FRB}~(\mbox{400--800}\,MHz) and FAST~(\mbox{1.0--1.5}\,GHz) observations would probe different parts of the emission spectrum. We therefore regard the FAST~results as inconclusive rather than as strong evidence against the \mbox{36.4-day} candidate activity period.

A candidate periodicity of~$\sim$200\,days has also been reported for FRB~20220529A, based on the temporal evolution of burst~RMs detected during FAST~monitoring observations~\citep{Liang+2025}. This period does not appear to be related to the \mbox{36.4-day} candidate activity period, either harmonically or through sidereal aliasing (see Section~\ref{sec:activity_period_aliasing}). If both candidate periodicities are real, they may therefore arise from distinct physical processes: one modulating the burst activity, and the other producing variations in the magneto-ionic environment along the line of sight. Continued monitoring of both the burst rate and the~RM of this source will enable further tests of this possibility.


\section{Discussion}
\label{sec:discussion}

In Section~\ref{sec:results}, we identified a new candidate activity period for FRB~20240209A, along with very tentative evidence for an activity period derivative. We also reported candidate activity periods for three repeating~FRB sources in \mbox{CHIME/FRB} Catalog~2: FRB~20190208A, FRB~20200926A, and FRB~20220529A. In this section, we discuss the astrophysical implications of these candidate periodicities.


\subsection{Revisiting the Candidate Activity Period of FRB~20240209A}
\label{sec:frb20240209a_activity_period_revisited}

\citet{Pal2025} reported an activity period of 128.5\,$\pm$\,6\,days for FRB~20240209A, based on $N_{\mathrm{Pal}}$\,$=$\,64~bursts detected with \mbox{CHIME/FRB} over $T_{\mathrm{Pal}}$\,$=$\,1.1\,years. Our substantially larger sample consists of $N$\,$=$\,155~bursts spanning $T$\,$=$\,3.5\,years (see Table~\ref{table:frb20240209a_burst_log}). If the activity period does not evolve over time, the \mbox{128.5-day} period is inconsistent with our \mbox{170.4-day} candidate activity period at the 5.5$\sigma$~level~(see Section~\ref{sec:frb20240209a_results}). The ratio of the two periods~(1.33\,$\pm$\,0.07) indicates that they are unlikely to be harmonically related.

The two periods could, however, be reconciled if the activity period evolves in time. For a constant activity period derivative, the activity period at time~$t$ is given by a first-order Taylor expansion about a reference time,~$t_{\mathrm{ref}}$:
\begin{equation}
	P(t) = P_{0} + \dot{P}\,(t - t_{\mathrm{ref}}),
	\label{eqn:activity_period_time_evolution}
\end{equation}
where $P_{0}$\,$=$\,$P(t_{\mathrm{ref}})$ is the activity period at the reference time and $\dot{P}$ is the activity period derivative. Using the values from Section~\ref{sec:frb20240209a_results} ($P_{0}$\,$=$\,166.0$^{\mathrm{+4.0}}_{\mathrm{-3.8}}$\,days, $\dot{P}$\,$=$\,0.20$^{\mathrm{+0.04}}_{\mathrm{-0.04}}$\,day\,day$^{\mathrm{-1}}$, and $t_{\mathrm{ref}}$\,$=$\,MJD\,60712.47454), Equation~(\ref{eqn:activity_period_time_evolution}) predicts an activity period of 131.0$^{\mathrm{+8.4}}_{\mathrm{-8.3}}$\,days at the mid-time~(MJD\,60541.99474) of the bursts analyzed in~\citet{Pal2025}. If our tentative~$\dot{P}$ is genuine, the two measurements could therefore reflect the same evolving periodicity sampled at different epochs. However, this scenario requires the activity period to change by nearly a factor of two over the 1.1\,years spanned by the bursts in~\citet{Pal2025}, an extreme rate of evolution not observed in any other periodic FRB~source. For this reason, and given the caveats discussed in Section~\ref{sec:frb20240209a_results}, we regard the candidate~$\dot{P}$ for FRB~20240209A as very tentative.

This comparison also illustrates a broader challenge in periodicity searches of sparse FRB~burst arrival times. Applying our FRB~periodicity search pipeline to the 64~bursts analyzed in~\citet{Pal2025}, we find a peak at 122.8\,$\pm$\,4.3\,days, 122.8\,$\pm$\,3.7\,days, and 122.8\,$\pm$\,2.6\,days in the~LS, $Z_{1}^{2}$, and \mbox{$H$-test} periodograms, with trials-corrected statistical significances of 7.4$\sigma$, 9.1$\sigma$, and 7.5$\sigma$, respectively. Although these measurements are consistent with the \mbox{128.5-day} period reported by~\citet{Pal2025} to within 1$\sigma$, no corresponding peak appears in the periodograms computed from all 155~bursts in Table~\ref{table:frb20240209a_burst_log} (see Figure~\ref{fig:frb20240209a_periodogram_activity_profile}\mbox{a--c}), indicating that this feature is present in the earlier data but does not persist as the burst sample grows. This behavior can arise because activity periods inferred from bursts spanning only a few cycles are sensitive to cycle-to-cycle variations in the source's burst activity. Confirmation of such measurements requires larger burst samples spanning more activity cycles, with significances evaluated against null models that account for these effects~\citep{Pearlman+2026a}.


\subsection{Physical Origins and Progenitor Scenarios for FRB~20240209A}
\label{sec:frb20240209a_physical_origins}

The \mbox{170.4-day} candidate activity period of FRB~20240209A, together with the source's unusual host environment, motivates a re-examination of its physical origins. The source was localized to the extreme outskirts of a luminous, quiescent elliptical galaxy at a redshift of $z$\,$=$\,0.1384, with a projected physical offset of 40\,$\pm$\,5\,kpc from the host center~\citep{Eftekhari+2025, Shah+2025}. This corresponds to a host-normalized offset of 5.1$R_{\mathrm{eff}}$, among the largest measured for any well-localized~FRB and exceeded only by that of FRB~20200120E, which resides in a globular cluster~(GC) offset from the M81~galaxy~\citep{Bhardwaj+2021a, Kirsten+2022, Pearlman+2025a}. The host's old stellar population (mass-weighted age of $\sim$11\,Gyr;~\citealt{Eftekhari+2025}) strongly disfavors a young magnetar formed in a recent core-collapse supernova.

Two progenitor scenarios appear more plausible. The first is a GC~association: the spatial distribution of~GCs in massive elliptical galaxies is consistent with the observed offset, and GC~environments can accommodate delayed formation channels, including accretion-induced collapse of a white dwarf and double white-dwarf or neutron-star mergers~\citep{Kremer+2021b, Kirsten+2022, Lu+2022}. Alternatively, the~FRB may originate in an as-yet-undetected satellite dwarf galaxy within the host's halo~\citep{Shah+2025}. This possibility is further supported by observations of other repeating~FRBs that have been localized to low-mass, faint host galaxies. For instance, \citet{Hewitt+2024b} identified an exceptionally low-luminosity dwarf host galaxy for FRB~20190208A, while \citet{Bhardwaj+2025}~found that FRB~20240114A resides in a low-metallicity dwarf satellite of a more massive central galaxy, analogous to the Small Magellanic \mbox{Cloud--Milky} Way system. In either scenario, the large projected offset disfavors an in situ origin within the main stellar body of the elliptical galaxy. However, the two possibilities imply distinct progenitor environments: a~GC host would point to an old, dynamically evolved stellar population, whereas a satellite dwarf galaxy could sustain more recent star formation.

The \mbox{170.4-day} candidate activity period can be interpreted within the principal classes of models proposed to explain long-term periodicity in repeating~FRBs. These models fall broadly into three categories: (i)~binary orbital models, in which the activity window is governed by the geometry of the companion's wind or an accretion disk~\citep{Ioka+2020, Lyutikov+2020a, Deng+2021, Sridhar+2021, Wada+2021, Barkov+2022}; (ii)~precession models, in which the emission beam periodically sweeps across the line of sight as the compact object precesses~\citep{Levin+2020, Tong+2020, Zanazzi+2020, Wasserman+2022}; and (iii)~ultra-long-period rotation models, in which the activity period corresponds to the neutron star's spin period~\citep{Beniamini+2020a}. The last scenario has also been invoked to explain ultra-long-period radio transients~\citep{Beniamini+2023}, such as \mbox{GLEAM-X}~J162759.5$-$523504~\citep{Hurley-Walker+2022}.

If the \mbox{170.4-day} candidate period is the orbital period of a neutron-star binary with a low-mass companion, significant mass transfer onto the neutron star would require an evolved donor. For a low-mass~(\mbox{0.3--1}\,$M_{\odot}$) companion and a canonical~(1.4\,$M_{\odot}$) neutron star, Kepler's Third Law yields a semi-major axis of $a$\,$\approx$\,\mbox{0.7--0.8}\,AU. At this separation, the companion's Roche lobe has a radius of $R_{\mathrm{L}}$\,$\approx$\,\mbox{40--60}\,$R_{\odot}$~\citep{Eggleton1983}. An unevolved low-mass star would be far smaller than its Roche lobe, and its wind would be too weak to supply significant accretion across such a wide orbit, so mass transfer through either Roche-lobe overflow or wind accretion would require a red-giant donor, consistent with a low-mass X-ray binary~(LMXB) hosting a giant companion. Neutron-star binaries with comparable orbital periods have been found in Galactic~GCs. For example, PSR~B1620$-$26 in~M4~(NGC~6121) has a \mbox{191-day} orbit with a $\sim$0.3\,$M_{\odot}$ white-dwarf companion, the expected remnant of mass transfer from a giant donor~\citep{Lyne+1988a, Sigurdsson+2003}. An evolved donor is thus naturally accommodated by environments with an old stellar population. By contrast, a massive \mbox{O/B-star} companion, such as those invoked in binary models proposed for FRB~20180916B~\citep{Ioka+2020, Lyutikov+2020a, Deng+2021}, would be difficult to accommodate in a~GC, where such short-lived stars are no longer present. If the source instead resides in a satellite galaxy with recent star formation, a massive companion would be plausible.

A \mbox{170.4-day} periodicity could also arise in a binary system through periodic changes in the viewing geometry or local propagation environment. One proposed scenario involves an accreting~LMXB or ultraluminous \mbox{X-ray} source~(ULX), where short-lived relativistic outflows from the compact object produce~FRBs via a synchrotron maser mechanism~\citep{Lyubarsky2014, Beloborodov2017, Waxman2017, Metzger+2019}, with the long-term periodicity driven by \mbox{Lense--Thirring} precession of the accretion disk and jet~\citep{Sridhar+2021}. The overabundance of~LMXBs and the presence of~ULXs in~GCs~\citep{Clark1975, Maccarone+2007a, Dage+2021} suggest that an accreting-binary origin is possible for FRB~20240209A. However, previous \mbox{X-ray} non-detections of the source exclude only~ULXs with $L_{\mathrm{X}}$\,$\gtrsim$\,10$^{\mathrm{43}}$\,erg\,s$^{\mathrm{-1}}$~\citep{Verrecchia+2025}. At the luminosity distance of FRB~20240209A's host galaxy~($d_{\mathrm{L}}$\,$\approx$\,675\,Mpc), deep \mbox{X-ray} observations reaching fluxes of $F_{\mathrm{X}}$\,$\approx$\,10$^{\mathrm{-15}}$\,erg\,s$^{\mathrm{-1}}$\,cm$^{\mathrm{-2}}$~($L_{\mathrm{X}}$\,$\approx$\,5\,$\times$\,10$^{\mathrm{40}}$\,erg\,s$^{\mathrm{-1}}$) would probe a greater range of ULXs than existing limits, although typical~LMXBs~($L_{\mathrm{X}}$\,$\lesssim$\,10$^{\mathrm{38}}$\,erg\,s$^{\mathrm{-1}}$) would remain undetectable~\citep{Pearlman+2025a}.

Non-orbital scenarios may also explain the candidate \mbox{170.4-day} periodicity. In precession models, the activity period corresponds to the precession period of the neutron star's spin axis, which arises either from free precession of a magnetically deformed neutron star~\citep{Levin+2020, Zanazzi+2020, Wasserman+2022} or from forced precession driven by an external torque, such as that exerted by a fallback disk~\citep{Tong+2020}. Alternatively, in ultra-long-period rotation models, mass-loaded particle winds or fallback accretion can spin down a neutron star born with a strong internal magnetic field to such long periods, while it retains a magnetar-strength field that could still power FRB~emission~\citep{Beniamini+2020a}.

The candidate activity period derivative of FRB~20240209A, if genuine, is difficult to explain physically. For example, if the \mbox{170.4-day} activity period is the orbital period of a binary, Kepler's Third Law implies an orbital expansion rate of $\dot{a} / a$\,$\approx$\,0.29\,year$^{\mathrm{-1}}$ for a system with approximately constant total mass. At this rate, the orbital separation would double in only $\sim$4\,years, far faster than can be explained by standard binary evolution~\citep{Postnov+2014}. Interpreting the candidate activity period and its derivative as the rotational period and spin-down rate of a neutron star is also implausible. Under vacuum magnetic dipole braking, this spin-down would imply a surface magnetic field of $B_{\mathrm{S}}$\,$\approx$\,5\,$\times$\,10$^{\mathrm{22}}$\,G, far above the $B_{\mathrm{max}}$\,$\approx$\,10$^{\mathrm{18}}$\,G virial limit for a neutron star~\citep{Lai+1991}. Neither fallback accretion torques nor a mass-loaded particle wind can supply the additional spin-down without similarly extreme conditions. Accretion torques would require a surface magnetic field of $B_{\mathrm{S}}$\,$\gtrsim$\,10$^{\mathrm{17}}$\,G for the inflow to couple to the magnetosphere near the corotation radius~($R_{\mathrm{co}}$\,$\approx$\,0.7\,AU), while a~particle wind from a 10$^{\mathrm{15}}$\,G magnetar would instead need a luminosity of $L_{\mathrm{pw}}$\,$\approx$\,5\,$\times$\,10$^{\mathrm{39}}$\,erg\,s$^{\mathrm{-1}}$, enough to drain the magnetar's entire external magnetic energy within a few years~\citep{Harding+1999, Beniamini+2020a}. We therefore consider it more likely that the candidate~$\dot{P}$ reflects cycle-to-cycle variations in the source's bursting activity~(see Section~\ref{sec:frb20240209a_results}).

The tentative secular decline in the~DM of FRB~20240209A~(\citealt{Cui+2026}; V.~Shah et al., in preparation) shares similarities with the decline seen in FRB~20220529A~\citep{Pandhi+2026} and may indicate an evolving local environment~\citep{Yang+2017, Piro+2018}. If both the candidate activity period and the DM~decline are confirmed, a viable model must explain both, for example through ionized material from a binary companion, either carried by its wind or ablated by the neutron star. Continued monitoring of the source's activity and~DM will be essential to confirm both features and determine whether they share a common physical origin.


\subsection{An Emerging Population of Periodic Repeating FRBs?}
\label{sec:periodic_repeating_frb_population}

The clearest periodicity detected from a repeating~FRB source remains that of FRB~20180916B~\citep{CHIME+2020b, Sand+2023, Pearlman+2026a}, which has an activity period of $P$\,$=$\,16.335\,$\pm$\,0.008\,days and no statistically significant activity period derivative ($\dot{P}$\,$=$\,$-$1.3$^{\mathrm{+4.0}}_{\mathrm{-4.0}}$\,$\times$\,10$^{\mathrm{-5}}$\,day\,day$^{\mathrm{-1}}$). This source exhibits a $\sim$\mbox{4--5-day} activity window, where the peak activity phase is chromatic~\citep{Pleunis+2021b, Pastor-Marazuela+2021, Bethapudi+2023, Gopinath+2024}. In the \mbox{400--800}\,MHz band, the burst rate is consistent with a constant-rate Poisson process throughout the activity window~\citep{Sand+2023}. Although \citet{Sand+2023} found no evidence of DM~variations over a span of $\sim$3\,years, \citet{Mckinven+2023a} identified a sudden, sustained linear increase in the~RM lasting $\sim$1\,year, after which the RM began varying stochastically around a nearly constant value of~$-$58.75\,rad\,m$^{\mathrm{-2}}$~\citep{Bethapudi+2025, Ng+2025b}.

The \mbox{170.4-day} candidate activity period of FRB~20240209A is substantially longer than the \mbox{16.3-day} activity period of FRB~20180916B, but it is comparable to the tentative \mbox{160-day} periodicity reported for FRB~20121102A~\citep{Rajwade+2020b, Cruces+2021, Braga+2025, Gouiffes+2025, Wang+2025}. FRB~20240209A and FRB~20121102A both exhibit large duty cycles ($\delta$\,$=$\,0.89 and $\delta$\,$\approx$\,0.6, respectively; see Section~\ref{sec:frb20240209a_results} and~\citealt{Rajwade+2020b}), whereas FRB~20180916B has a smaller duty cycle of $\delta$\,$\approx$\,0.3~\citep{Sand+2023, Pearlman+2026a}. We note that the measured duty cycle of FRB~20180916B is frequency-dependent, with the activity window narrowing toward higher frequencies~\citep{Pleunis+2021b, Pastor-Marazuela+2021, Bethapudi+2023, Gopinath+2024}. For FRB~20121102A, strong burst clustering, an incomplete exposure function that does not fully account for non-detections, and irregularly spaced observations from multiple telescopes lead to non-uniform phase sampling, complicating assessments of the statistical significance of the \mbox{160-day} periodicity. FRB~20180916B is therefore currently the only repeater with a well-established long-term periodicity in its burst activity. Together with FRB~20121102A, the candidate periodicities we report for FRB~20240209A, FRB~20190208A, FRB~20200926A, and FRB~20220529A bring the total number of confirmed and candidate periodic repeaters to six.

The candidate periodic repeaters exhibit a wide range of apparent activity periods~($P$\,$\approx$\,\mbox{28--170}\,days), duty cycles~($\delta$\,$\approx$\,\mbox{0.16--0.89}), and degrees of regularity in their burst activity. If these candidate periodicities are confirmed, their diversity would raise the possibility that the mechanism responsible for FRB~20180916B's \mbox{16.3-day} periodicity differs from those operating in other repeaters. The activity period of FRB~20180916B is measured to a fractional precision of $\sigma_{P}/P$\,$\approx$\,5\,$\times$\,10$^{\mathrm{-4}}$, and its activity period derivative is consistent with zero~\citep{Sand+2023, Pearlman+2026a}. This level of stability points to a well-defined dynamical process, such as orbital motion in a binary system~\citep{Ioka+2020, Lyutikov+2020a}, or the rotation~\citep{Beniamini+2020a} or precession of a compact object~\citep{Levin+2020, Tong+2020, Zanazzi+2020, Wasserman+2022}. In contrast, the candidate activity periods of FRB~20240209A and FRB~20121102A are far less precisely constrained, and their bursts are spread over a much wider range of activity phases. 

The long periods and wide activity windows observed from some repeaters may instead reflect a characteristic timescale of activity rather than a strict periodicity. In this case, the physically relevant timescale is not the period itself but the effective activity window, $W_{\mathrm{eff}}$\,$=$\,$\delta$\,$\times$\,$P_{\mathrm{activity}}$, from Equation~(\ref{eqn:duty_cycle}). For the candidate periodic sources identified in this work, we find $W_{\mathrm{eff}}$\,$\approx$\,18, 14, and 22\,days for FRB~20190208A, FRB~20200926A, and FRB~20220529A, respectively, and $W_{\mathrm{eff}}$\,$\approx$\,151\,days for FRB~20240209A, whose bursts are concentrated in an enhanced activity window lasting $\sim$72\,days. These durations fall within the range spanned by Galactic magnetar outbursts, which last from a few weeks to years as magnetic energy stored in the crust and the magnetosphere is released~\citep{Kaspi+2017, CotiZelati+2018}. If~FRBs are produced by magnetars, such outbursts could set the duration of burst activity, with crustal stresses building up as the internal magnetic field evolves~(e.g.,~\citealt{Perna+2011}). The Galactic magnetar SGR~1935$+$2154, which emitted an FRB-like radio burst during one of its active episodes~\citep{CHIME+2020a, Bochenek+2020c}, provides a direct analog. Its high-energy burst activity has been reported to recur in episodes consistent with a window period of 231\,$\pm$\,9\,days and a duty cycle of $\delta$\,$\approx$\,0.55~($W_{\mathrm{eff}}$\,$\approx$\,130\,days), based on six active episodes~\citep{Grossan2021, Denissenya+2021}. Both the period and the effective activity window are comparable to those of FRB~20240209A and FRB~20121102A. Such energy-driven activity would be quasi-periodic, rather than strictly periodic, so the activity phase would drift over many cycles, a prediction that continued monitoring can test directly.

Periodicity searches of \mbox{CHIME/FRB} burst arrival times from repeating~FRB sources are not uniformly sensitive to all activity periods. Since \mbox{CHIME/FRB} is a transit instrument, the shortest detectable period is nominally $\sim$2\,days. In practice, however, periods of $\sim$\mbox{2--10}\,days are often difficult to distinguish from aliases generated by the sidereal frequency, so 10\,days is a more practical lower limit. FRB~20180916B's \mbox{16.3-day} activity period is the most statistically significant periodicity detected to date, in part because of its short period (completing 22.4\,cycles\,year$^{\mathrm{-1}}$), small duty cycle, and sustained activity over many cycles. In contrast, the \mbox{170.4-day} candidate activity period of FRB~20240209A completes only 2.1\,cycles\,year$^{\mathrm{-1}}$, and the source has the largest duty cycle among the confirmed and candidate periodic repeaters, making its periodicity substantially more difficult to detect. In sources with large duty cycles, bursts may occur throughout the activity cycle at a low rate, with certain phases exhibiting substantially enhanced activity (e.g.,~Figure~\ref{fig:frb20240209a_periodogram_activity_profile}\mbox{d--e} for FRB~20240209A). These selection effects underscore the importance of sustained, sensitive, wide-field daily monitoring of repeating~FRB sources over long observing baselines, as uniquely provided by~\mbox{CHIME/FRB}. Drawing robust statistical conclusions about the underlying distribution of activity periods in the repeating~FRB population will require a substantially larger sample of periodic sources.

Finally, the local environments of the three repeaters with the strongest evidence for periodic activity~(FRB~20180916B, FRB~20121102A, and FRB~20240209A) are remarkably diverse. FRB~20180916B resides in the massive star-forming galaxy SDSS~J015800.28$+$654253.0~\citep{Marcote+2020}, with a projected offset of $\sim$250\,pc from the nearest star-forming region~\citep{Tendulkar+2021}. Based on this offset, \citet{Tendulkar+2021} argue that FRB~20180916B may be a neutron star in a binary with a massive star, such as a high-mass \mbox{X-ray} binary or a gamma-ray binary with a late \mbox{O/B-type} companion. FRB~20121102A lies within a star-forming region of a dwarf galaxy~\citep{Bassa+2017, Tendulkar+2017} and is associated with a luminous, compact persistent radio source~\citep{Chatterjee+2017, Marcote+2017}, whereas FRB~20240209A is located in the far outskirts of a quiescent elliptical galaxy~\citep{Eftekhari+2025, Shah+2025}. Of the other three candidate periodic repeaters identified in this work, only two have known host galaxies: FRB~20190208A has been localized to an extremely low-luminosity dwarf galaxy~\citep{Hewitt+2024b}, and the host of FRB~20220529A is a disk galaxy at a redshift of $z$\,$=$\,0.18~\citep{Li+2026a}. Taken together, these markedly different environments do not point to an obvious common progenitor channel, even among the sources with the strongest evidence for periodic activity.


\subsection{Aliasing and Confirmation of the Candidate Activity Periods}
\label{sec:activity_period_aliasing}

Since \mbox{CHIME/FRB} is a transit instrument that observes each source for only short intervals each day, its regular daily sampling can introduce a degeneracy between the frequency of the activity period~($f$\,$=$\,$1/P$) and a family of aliases,
\begin{equation}
	f_{\mathrm{alias}} = |f \pm (m \times f_{\mathrm{sid}})|,
	\label{eqn:alias_frequencies}
\end{equation}
where $m$ is a non-negative integer denoting the alias order and $f_{\mathrm{sid}}$\,$=$\,1/(0.99727\,days) is the sidereal frequency.

\citet{CHIME+2020b} presented a heuristic argument that bounds~$m$ for FRB~20180916B, based on the requirement that the duration of the activity window exceed the time that the source is observed during each transit. The resulting bound, $m$\,$\lesssim$\,$\delta/(t_{\mathrm{exp}}\,f_{\mathrm{sid}})$, where $t_{\mathrm{exp}}$ is the source's exposure time per transit, is restrictive only when the bursts are tightly clustered in activity phase. Applying the same argument to FRB~20240209A, FRB~20190208A, FRB~20200926A, and FRB~20220529A, we find that a broad range of~$m$ values remains permitted, so the alias order is poorly constrained for the candidate periodic sources identified in this work.

Given the fixed daily cadence of \mbox{CHIME/FRB}, we cannot exclude, using \mbox{CHIME/FRB}~ToAs alone, the possibility that the candidate activity periods reported here are sidereal aliases of a different underlying periodicity. We emphasize, however, that our null model already incorporates \mbox{CHIME/FRB's} daily sampling. The observing cadence alone therefore cannot produce a statistically significant signal, and any such signal, even if it appears at an alias frequency, would require genuine astrophysical structure in the burst arrival times. Bursts detected outside the \mbox{CHIME/FRB}~transit windows would directly test whether the candidate activity periods reflect an intrinsic periodicity of the source. We therefore strongly encourage complementary observations of these sources with other radio telescopes at similar observing frequencies.

The FAST~observations of FRB~20220529A presented in Section~\ref{sec:frb20220529a_results} provide a first test of this kind, but the result is inconclusive: the FAST~bursts were detected at higher frequencies~(\mbox{1.0--1.5}\,GHz), where the periodic activity may differ, and most were detected within only $\sim$7~consecutive activity cycles~\citep{Li+2026a}. Similar tests for FRB~20190208A, FRB~20200926A, and FRB~20240209A will require sustained monitoring with telescopes that observe each source at different times than \mbox{CHIME/FRB}, which can help break the degeneracy described by Equation~(\ref{eqn:alias_frequencies}). Continued monitoring of the repeating~FRB sources shown in Figure~\ref{fig:frb_toa_timeline}, especially those with candidate activity periods~(see Table~\ref{table:frb_periodicity_search_summary}), will be essential for determining whether these periodicities are intrinsic.


\section{Conclusions and Future Prospects}
\label{sec:conclusions}

Using a new FRB~periodicity search pipeline introduced by~\citet{Pearlman+2026a}, we searched for periodic activity in the \mbox{CHIME/FRB}~burst arrival times from FRB~20240209A and 19~repeating~FRB sources in \mbox{CHIME/FRB} Catalog~2~\citep{CHIME+2026, Cook+2026}. Our main findings, summarized in Table~\ref{table:frb_periodicity_search_summary}, are as follows:
\begin{itemize}
	
	\item For FRB~20240209A, we report a new candidate activity period of 170.4\,$\pm$\,4.7\,days, detected with a trials-corrected statistical significance of 5.3$\sigma$ (see Section~\ref{sec:frb20240209a_results}). We also find very tentative evidence for an associated activity period derivative~($\dot{P}$\,$=$\,0.20$^{\mathrm{+0.04}}_{\mathrm{-0.04}}$\,day\,day$^{\mathrm{-1}}$). Our larger burst sample, which spans a substantially longer observing baseline, shows no evidence for the \mbox{128.5-day} activity period previously reported for this source by~\citet{Pal2025}.
	
	\item The burst arrival times from FRB~20190208A show a candidate activity period of 28.46\,$\pm$\,0.08\,days, with a trials-corrected statistical significance of 3.1$\sigma$ (see Section~\ref{sec:frb20190208a_results}).
	
	\item At a trials-corrected statistical significance of 4.0$\sigma$, we identify a candidate activity period of 86.7\,$\pm$\,0.5\,days for FRB~20200926A (see Section~\ref{sec:frb20200926a_results}).
	
	\item FRB~20220529A exhibits a candidate activity period of 36.4\,$\pm$\,0.5\,days, which we detect with a trials-corrected statistical significance of 3.5$\sigma$ (see Section~\ref{sec:frb20220529a_results}).
	
\end{itemize}
These significances are evaluated against a homogeneous (constant-rate) Poisson null hypothesis and may decrease if burst clustering is taken into account (see Section~\ref{sec:frb_periodicity_search_pipeline_overview}), so these candidate activity periods should be regarded as provisional.

As a guideline, we would consider a candidate activity period to be robustly established if it reached a trials-corrected statistical significance of $\gtrsim$\,5$\sigma$ against a null model that accounts for burst clustering and remained stable over many independent activity cycles, as has been demonstrated for FRB~20180916B~\citep{CHIME+2020b, Sand+2023, Pearlman+2026a}. Meeting these criteria for the four sources with candidate activity periods will require additional bursts spanning more activity cycles. Given their periods and burst rates, this will likely take years of continued monitoring, and the required amount of data will vary substantially from source to source, depending on the burst rate, degree of clustering, and duty cycle.

More broadly, we find that periodic activity is not readily detectable across the population of \mbox{CHIME/FRB}~repeaters. Of the 20~sources we searched, only FRB~20180916B has a firmly established activity period. It remains unclear whether periodic activity is intrinsically rare among repeaters or difficult to detect with current data. Resolving this question will require sustained, regular monitoring of many repeaters.

\mbox{CHIME/FRB} is well suited to such monitoring and is likely to remain the premier instrument for discovering long-timescale (days or longer) periodic activity in repeating~FRBs for the foreseeable future. Its broad observing bandwidth is well matched to repeater bursts, which often occupy a limited fraction of the observing band and exhibit substantial spectral variability~\citep{Pleunis+2021a}. Additionally, \mbox{CHIME/FRB's} large field of view and near-daily cadence allow it to sample the activity cycles of many sources simultaneously with roughly uniform exposure, a capability that is essential for distinguishing between periodically modulated and purely stochastic activity.

Upcoming radio telescopes with higher sensitivity, such as the Canadian Hydrogen Observatory and Radio-transient Detector~(CHORD;~\citealt{Vanderlinde+2019, CHORD+2026}) and the Deep Synoptic Array~(DSA;~\citealt{Hallinan+2019}), will have much smaller fields of view and therefore lack the broad daily sky exposure that enables \mbox{CHIME/FRB} to monitor large numbers of repeating~FRB sources regularly. Some pointed instruments, such as FAST and the Green Bank Telescope~(GBT), are substantially more sensitive than \mbox{CHIME/FRB}, but they are oversubscribed, shared facilities that cannot easily provide long-term monitoring at a regular cadence. They nonetheless remain essential for the off-transit detections needed to resolve sidereal aliasing (see Section~\ref{sec:activity_period_aliasing}). At the other extreme, ultra-wide-field detectors, such as the Bustling Universe Radio Survey Telescope~(BURSTT;~\citealt{Lin+2022}) and the \mbox{256-antenna} Coherent \mbox{All-Sky} Monitor~(\mbox{CASM-256};~\citealt{Connor+2026}), will generally lack the sensitivity required to accumulate the burst counts needed to identify periodicity robustly. The upcoming \mbox{CHIME/FRB}~\mbox{X-engine} and \mbox{FRB-compute} upgrades are together expected to increase the FRB~detection rate by a factor of $\gtrsim$\,3 by improving coverage of each source during its transit through CHIME's primary beam. This higher detection rate will substantially accelerate both the discovery and confirmation of periodicities in the repeating~FRB population.


\begin{acknowledgments}

We thank Barak Zackay for useful discussions.

A.B.P.~acknowledges support from NASA through the NASA Hubble Fellowship grant \mbox{HST-HF2-51584.001-A} awarded by the Space Telescope Science Institute, which is operated by the Association of Universities for Research in Astronomy, Inc., for NASA, under contract \mbox{NAS5-26555}. A.B.P.~also acknowledges prior support from a Banting Fellowship, a McGill Space Institute~(MSI) Fellowship, and a Fonds de Recherche du \mbox{Qu\'ebec -- Nature} et Technologies~(FRQNT) Postdoctoral Fellowship. A.M.C.~is a Banting Postdoctoral Researcher. V.M.K.~holds the Lorne Trottier Chair in Astrophysics \& Cosmology, a Distinguished James McGill Professorship, and receives support from a Natural Sciences and Engineering Research Council of Canada~(NSERC) Discovery Grant~(RGPIN~\mbox{228738-13}). K.W.M.~is supported by National Science Foundation~(NSF) Grant No.~\mbox{AST-2510771}. A.P.C.~is a Canadian Square Kilometre Array~(SKA) Scientist and is funded by the Government of Canada / est financ\'e par le gouvernement du Canada. M.D.~is supported by a Canada Research Chair~(CRC), an NSERC Discovery Grant, and the Canadian Institute for Advanced Research~(CIFAR). E.F.~is supported by~NSF Grant No.~\mbox{AST-2407399}. A.P.~is a Trottier Space Institute~(TSI) Postdoctoral Fellow. M.W.S.~acknowledges support from the French government under the France~2030 investment plan, as part of the Initiative d'Excellence \mbox{d'Aix-Marseille} \mbox{Universit\'e -- A*MIDEX}~\mbox{(AMX-23-CEI-088).} P.S.~acknowledges support from an NSERC~Discovery Grant~\mbox{(RGPIN-2024-06266).} V.S.~is supported by an FRQNT~Doctoral Research Award.

This research was enabled in part by support provided by the British Columbia Digital Research Infrastructure Group and the Digital Research Alliance of Canada~(\href{https://www.alliancecan.ca}{alliancecan.ca}).

The AstroFlash research group at McGill University, the University of Amsterdam, ASTRON, and~JIVE is supported by: a Canada Excellence Research Chair~(CERC) in Transient Astrophysics~\mbox{(CERC-2022-00009);} an Advanced Grant from the European Research Council~(ERC) under the European Union's Horizon~2020 research and innovation programme (`EuroFlash': Grant agreement No.~101098079); an \mbox{NWO-Vici} grant (`AstroFlash': VI.C.192.045); an NSERC~Discovery Grant~\mbox{(RGPIN-2025-06681);} an ERC~Starting Grant (`EnviroFlash': Grant agreement No.~101223057); and an \mbox{NWO-Veni} grant (VI.Veni.222.295). Pulsar and FRB~research at the University of British Columbia is funded by an NSERC~Discovery Grant and by~CIFAR.

We acknowledge that CHIME is located on the traditional, ancestral, and unceded territory of the \mbox{Syilx/Okanagan} people. We are grateful to the staff of the Dominion Radio Astrophysical Observatory, which is operated by the National Research Council of Canada.

CHIME operations are funded by a grant from the NSERC~Alliance Program and by support from McGill University, the University of British Columbia, and the University of Toronto. CHIME~was funded by a grant from the Canada Foundation for Innovation~(CFI) 2012~Leading Edge Fund (Project~31170) and by contributions from the provinces of British Columbia, Qu\'ebec, and Ontario. The \mbox{CHIME/FRB}~Project was funded by a grant from the CFI~2015 Innovation Fund~(Project~33213), by contributions from the provinces of British Columbia and Qu\'ebec, and by the Dunlap Institute for Astronomy and Astrophysics at the University of Toronto. Additional support was provided by CIFAR, the TSI~at McGill University, and the University of British Columbia. The \mbox{CHIME/FRB}~baseband recording system is funded in part by a CFI~John~R.~Evans Leaders Fund award to~I.H.S.

\end{acknowledgments}


\clearpage

\startlongtable
\begin{deluxetable*}{ccccc}
	\tabletypesize{\small}
	\tablewidth{0pt}
	\setlength{\tabcolsep}{0.5cm}
	\tablecaption{Radio Bursts from FRB~20240209A}
	\tablehead{\\[-15pt]\colhead{Burst~ID} & \colhead{TNS Name\,\tablenotemark{a}} & \colhead{Time of Arrival\,\tablenotemark{b}} & \colhead{S/N\,\tablenotemark{c}} & \colhead{Dispersion Measure\,\tablenotemark{d}} \\
		\colhead{} & \colhead{} & \colhead{(ToA, MJD)} & \colhead{} & \colhead{(DM, pc\,cm$^{\mathrm{-3}}$)}}
	\startdata
	B1 & FRB 20220622C & 59752.30110 & 12.8 & 183.2\,$\pm$\,0.4 \\
	B2 & FRB 20240209A & 60349.29878 & 16.0 & 179.5\,$\pm$\,1.6 \\
	B3 & FRB 20240217A & 60357.27506 & 9.2 & 178.3\,$\pm$\,0.4 \\
	B4 & FRB 20240301A & 60370.11098 & 9.5 & 179.5\,$\pm$\,1.6 \\
	B5 & FRB 20240309A & 60378.18724 & 16.8 & 175.9\,$\pm$\,0.4 \\
	B6 & FRB 20240608A & 60469.43274 & 9.0 & 175.9\,$\pm$\,0.4 \\
	B7 & FRB 20240612A & 60473.36883 & 14.4 & 178.7\,$\pm$\,0.8 \\
	B8 & FRB 20240612B & 60473.89815 & 10.9 & 179.5\,$\pm$\,1.6 \\
	B9 & FRB 20240615A & 60476.33047 & 11.2 & 179.5\,$\pm$\,1.6 \\
	B10 & FRB 20240616A & 60477.95300 & 13.7 & 175.1\,$\pm$\,0.4 \\
	B11 & FRB 20240619A & 60480.37334 & 10.5 & 175.9\,$\pm$\,0.4 \\
	B12 & FRB 20240619B & 60480.37752 & 10.7 & 175.5\,$\pm$\,0.8 \\
	B13 & FRB 20240619C & 60480.42432 & 10.3 & 177.1\,$\pm$\,0.8 \\
	B14 & FRB 20240620A & 60481.37213 & 8.5 & 177.1\,$\pm$\,0.8 \\
	B15 & FRB 20240620B & 60481.38526 & 8.5 & 177.1\,$\pm$\,0.8 \\
	B16 & FRB 20240621A & 60482.87565 & 9.0 & 178.7\,$\pm$\,0.8 \\
	B17 & FRB 20240625A & 60486.84758 & 12.3 & 175.5\,$\pm$\,0.8 \\
	B18 & FRB 20240628A & 60489.92039 & 9.6 & 177.1\,$\pm$\,0.8 \\
	B19 & FRB 20240629A & 60490.33236 & 16.2 & 177.1\,$\pm$\,0.8 \\
	B20 & FRB 20240629B & 60490.35610 & 20.7 & 175.1\,$\pm$\,0.4 \\
	B21 & FRB 20240629C & 60490.37355 & 19.5 & 175.1\,$\pm$\,0.4 \\
	B22 & FRB 20240629D & 60490.46322 & 9.7 & 172.3\,$\pm$\,0.8 \\
	B23 & FRB 20240629E & 60490.80318 & 16.1 & 177.1\,$\pm$\,0.8 \\
	B24 & FRB 20240630A & 60491.42055 & 8.2 & 175.5\,$\pm$\,0.8 \\
	B25 & FRB 20240702A & 60493.35841 & 9.3 & 178.7\,$\pm$\,0.8 \\
	B26 & FRB 20240702B & 60493.86552 & 8.2 & 175.5\,$\pm$\,0.8 \\
	B27 & FRB 20240716A & 60507.25991 & 8.6 & 173.9\,$\pm$\,0.8 \\
	B28 & FRB 20240720A & 60511.25581 & 8.3 & 175.1\,$\pm$\,0.4 \\
	B29 & FRB 20240724O & 60515.31156 & 8.4 & 178.7\,$\pm$\,0.8 \\
	B30 & FRB 20240731A & 60522.76754 & 8.1 & 182.8\,$\pm$\,1.6 \\
	B31 & FRB 20241012A & 60595.03707 & 9.3 & 175.1\,$\pm$\,0.4 \\
	B32 & FRB 20241012B & 60595.08339 & 8.7 & 175.1\,$\pm$\,0.4 \\
	B33 & FRB 20241013A & 60596.03567 & 10.8 & 175.1\,$\pm$\,0.4 \\
	B34 & FRB 20241013B & 60596.62852 & 13.2 & 176.3\,$\pm$\,1.6 \\
	B35 & FRB 20241015C & 60598.07967 & 16.6 & 175.1\,$\pm$\,0.4 \\
	B36 & FRB 20241016A & 60599.06539 & 9.5 & 175.9\,$\pm$\,0.4 \\
	B37 & FRB 20241016B & 60599.54440 & 11.8 & 177.1\,$\pm$\,0.8 \\
	B38 & FRB 20241017E & 60600.57542 & 10.6 & 175.5\,$\pm$\,0.8 \\
	B39 & FRB 20241019E & 60602.06475 & 9.3 & 175.9\,$\pm$\,0.4 \\
	B40 & FRB 20241020A & 60603.07061 & 8.3 & 174.3\,$\pm$\,0.4 \\
	B41 & FRB 20241021A & 60604.03069 & 8.6 & 174.3\,$\pm$\,0.4 \\
	B42 & FRB 20241022B & 60605.02820 & 14.0 & 177.1\,$\pm$\,0.8 \\
	B43 & FRB 20241022C & 60605.51662 & 13.4 & 175.5\,$\pm$\,0.8 \\
	B44 & FRB 20241024C & 60607.59732 & 15.2 & 175.9\,$\pm$\,0.4 \\
	B45 & FRB 20241024D & 60607.97031 & 14.1 & 175.5\,$\pm$\,0.8 \\
	B46 & FRB 20241025B & 60608.51718 & 12.4 & 174.3\,$\pm$\,0.4 \\
	B47 & FRB 20241026B & 60609.46890 & 17.4 & 177.1\,$\pm$\,0.8 \\
	B48 & FRB 20241026C & 60609.47856 & 9.3 & 175.1\,$\pm$\,0.4 \\
	B49 & FRB 20241026D & 60609.53795 & 16.7 & 175.5\,$\pm$\,0.8 \\
	B50 & FRB 20241027D & 60610.01311 & 9.5 & 175.5\,$\pm$\,0.8 \\
	B51 & FRB 20241027E & 60610.99158 & 8.9 & 175.5\,$\pm$\,0.8 \\
	B52 & FRB 20241028B & 60611.06949 & 10.9 & 174.3\,$\pm$\,0.4 \\
	B53 & FRB 20241028C & 60611.52817 & 15.4 & 175.5\,$\pm$\,0.8 \\
	B54 & FRB 20241029A & 60612.53866 & 15.8 & 175.9\,$\pm$\,0.4 \\
	B55 & FRB 20241030B & 60613.49597 & 15.8 & 175.5\,$\pm$\,0.8 \\
	B56 & FRB 20241030A & 60613.53558 & 14.7 & 176.7\,$\pm$\,0.4 \\
	B57 & FRB 20241030C & 60613.98894 & 16.9 & 176.3\,$\pm$\,1.6 \\
	B58 & FRB 20241031B & 60614.00957 & 9.4 & 174.3\,$\pm$\,0.4 \\
	B59 & FRB 20241031C & 60614.05788 & 8.4 & 176.3\,$\pm$\,1.6 \\
	B60 & FRB 20241031D & 60614.06958 & 11.6 & 175.1\,$\pm$\,0.4 \\
	B61 & FRB 20241031E & 60614.94965 & 9.2 & 176.3\,$\pm$\,1.6 \\
	B62 & FRB 20241101A & 60615.97263 & 8.6 & 177.1\,$\pm$\,0.8 \\
	B63 & FRB 20241102B & 60616.48338 & 9.4 & 175.5\,$\pm$\,0.8 \\
	B64 & FRB 20241103B & 60617.06651 & 10.1 & 174.3\,$\pm$\,0.4 \\
	B65 & FRB 20241110B & 60624.01366 & 10.9 & 179.5\,$\pm$\,1.6 \\
	B66 & FRB 20241117A & 60631.01228 & 8.4 & 177.1\,$\pm$\,0.8 \\
	B67 & FRB 20241119A & 60633.45407 & 10.5 & 175.5\,$\pm$\,0.8 \\
	B68 & FRB 20241122A & 60636.97612 & 8.3 & 180.4\,$\pm$\,0.8 \\
	B69 & FRB 20241124B & 60638.00570 & 10.7 & 175.5\,$\pm$\,0.8 \\
	B70 & FRB 20241126B & 60640.95174 & 8.1 & 174.3\,$\pm$\,0.4 \\
	B71 & FRB 20241209B & 60653.38578 & 13.6 & 175.5\,$\pm$\,0.8 \\
	B72 & FRB 20241216C & 60660.92418 & 8.2 & 177.1\,$\pm$\,0.8 \\
	B73 & FRB 20241230B & 60674.89779 & 8.2 & 176.7\,$\pm$\,0.4 \\
	B74 & FRB 20250109A & 60684.82451 & 12.3 & 176.3\,$\pm$\,1.6 \\
	B75 & FRB 20250123B & 60698.27965 & 10.2 & 174.3\,$\pm$\,0.4 \\
	B76 & FRB 20250125B & 60700.72282 & 8.1 & 173.9\,$\pm$\,0.8 \\
	B77 & FRB 20250126B & 60701.80062 & 10.8 & 173.1\,$\pm$\,1.6 \\
	B78 & FRB 20250128A & 60703.78636 & 10.0 & 176.3\,$\pm$\,1.6 \\
	B79 & FRB 20250205B & 60711.73458 & 8.8 & 175.5\,$\pm$\,0.8 \\
	B80 & FRB 20250205C & 60711.75392 & 8.1 & 174.3\,$\pm$\,0.4 \\
	B81 & FRB 20250211A & 60717.29961 & 8.8 & 174.3\,$\pm$\,0.4 \\
	B82 & FRB 20250219C & 60725.77893 & 14.4 & 173.5\,$\pm$\,0.4 \\
	B83 & FRB 20250225B & 60731.70538 & 8.2 & 176.3\,$\pm$\,1.6 \\
	B84 & FRB 20250228C & 60734.69069 & 12.3 & 175.5\,$\pm$\,0.8 \\
	B85 & FRB 20250321B & 60755.67248 & 11.8 & 174.3\,$\pm$\,0.4 \\
	B86 & FRB 20250323B & 60757.12134 & 13.2 & 177.5\,$\pm$\,0.4 \\
	B87 & FRB 20250330A & 60764.66931 & 8.5 & 174.3\,$\pm$\,0.4 \\
	B88 & FRB 20250331C & 60765.09672 & 13.5 & 174.3\,$\pm$\,0.4 \\
	B89 & FRB 20250408B & 60773.09874 & 8.0 & 175.1\,$\pm$\,0.4 \\
	B90 & FRB 20250410F & 60775.61745 & 16.2 & 176.3\,$\pm$\,1.6 \\
	B91 & FRB 20250412C & 60777.01825 & 15.4 & 173.1\,$\pm$\,1.6 \\
	B92 & FRB 20250412D & 60777.50856 & 17.3 & 173.9\,$\pm$\,0.8 \\
	B93 & FRB 20250413A & 60778.09841 & 10.9 & 175.1\,$\pm$\,0.4 \\
	B94 & FRB 20250415A & 60780.51314 & 13.3 & 176.3\,$\pm$\,1.6 \\
	B95 & FRB 20250416A & 60781.06638 & 11.7 & 174.3\,$\pm$\,0.4 \\
	B96 & FRB 20250426A & 60791.47197 & 9.9 & 174.3\,$\pm$\,0.4 \\
	B97 & FRB 20250501B & 60796.05967 & 8.5 & 173.5\,$\pm$\,0.4 \\
	B98 & FRB 20250501C & 60796.53859 & 12.4 & 174.3\,$\pm$\,0.4 \\
	B99 & FRB 20250502B & 60797.56447 & 8.0 & 173.9\,$\pm$\,0.8 \\
	B100 & FRB 20250503A & 60798.46705 & 13.8 & 174.3\,$\pm$\,0.4 \\
	B101 & FRB 20250504B & 60799.03698 & 11.2 & 178.7\,$\pm$\,0.8 \\
	B102 & FRB 20250504C & 60799.45981 & 10.0 & 175.1\,$\pm$\,0.4 \\
	B103 & FRB 20250507D & 60802.06409 & 12.9 & 176.3\,$\pm$\,1.6 \\
	B104 & FRB 20250508C & 60803.04439 & 8.4 & 176.3\,$\pm$\,1.6 \\
	B105 & FRB 20250508A & 60803.50714 & 15.2 & 175.5\,$\pm$\,0.8 \\
	B106 & FRB 20250510B & 60805.53013 & 9.4 & 173.9\,$\pm$\,0.8 \\
	B107 & FRB 20250511B & 60806.42566 & 13.8 & 173.1\,$\pm$\,1.6 \\
	B108 & FRB 20250511C & 60806.49575 & 8.4 & 173.5\,$\pm$\,0.4 \\
	B109 & FRB 20250512A & 60807.50646 & 8.8 & 173.1\,$\pm$\,1.6 \\
	B110 & FRB 20250513C & 60808.45657 & 20.7 & 176.3\,$\pm$\,1.6 \\
	B111 & FRB 20250513D & 60808.46208 & 10.0 & 176.3\,$\pm$\,1.6 \\
	B112 & FRB 20250513E & 60808.50266 & 13.7 & 175.5\,$\pm$\,0.8 \\
	B113 & FRB 20250515C & 60810.97946 & 12.1 & 173.9\,$\pm$\,0.8 \\
	B114 & FRB 20250518B & 60813.00329 & 10.2 & 176.3\,$\pm$\,1.6 \\
	B115 & FRB 20250520B & 60815.98688 & 17.0 & 175.5\,$\pm$\,0.8 \\
	B116 & FRB 20250522B & 60817.96345 & 11.3 & 176.7\,$\pm$\,0.4 \\
	B117 & FRB 20250530B & 60825.51190 & 10.8 & 176.3\,$\pm$\,1.6 \\
	B118 & FRB 20250601B & 60827.43164 & 8.6 & 172.3\,$\pm$\,0.8 \\
	B119 & FRB 20250603C & 60829.91870 & 9.2 & 174.3\,$\pm$\,0.4 \\
	B120 & FRB 20250604C & 60830.41387 & 14.0 & 175.5\,$\pm$\,0.8 \\
	B121 & FRB 20250604D & 60830.97249 & 12.3 & 172.7\,$\pm$\,0.4 \\
	B122 & FRB 20250607D & 60833.45102 & 15.7 & 172.3\,$\pm$\,0.8 \\
	B123 & FRB 20250616B & 60842.37668 & 9.7 & 173.9\,$\pm$\,0.8 \\
	B124 & FRB 20250628B & 60854.86059 & 17.0 & 173.5\,$\pm$\,0.4 \\
	B125 & FRB 20250710A & 60866.35446 & 15.7 & 173.9\,$\pm$\,0.8 \\
	B126 & FRB 20250711A & 60867.81270 & 14.5 & 176.3\,$\pm$\,1.6 \\
	B127 & FRB 20250715C & 60871.79783 & 10.6 & 175.5\,$\pm$\,0.8 \\
	B128 & FRB 20250723A & 60879.34031 & 9.8 & 179.5\,$\pm$\,1.6 \\
	B129 & FRB 20250805B & 60892.27159 & 9.8 & 176.7\,$\pm$\,0.4 \\
	B130 & FRB 20250809A & 60896.24454 & 8.0 & 177.1\,$\pm$\,0.8 \\
	B131 & FRB 20250819B & 60906.74239 & 11.1 & 174.3\,$\pm$\,0.4 \\
	B132 & FRB 20250927A & 60945.09704 & 10.6 & 173.5\,$\pm$\,0.4 \\
	B133 & FRB 20250929A & 60947.13154 & 14.4 & 175.5\,$\pm$\,0.8 \\
	B134 & FRB 20251001A & 60949.08551 & 17.1 & 177.1\,$\pm$\,0.8 \\
	B135 & FRB 20251004B & 60952.59980 & 10.4 & 175.5\,$\pm$\,0.8 \\
	B136 & FRB 20251005B & 60953.57103 & 11.4 & 175.5\,$\pm$\,0.8 \\
	B137 & FRB 20251008A & 60956.05694 & 9.8 & 174.3\,$\pm$\,0.4 \\
	B138 & FRB 20251010C & 60958.54459 & 14.3 & 176.3\,$\pm$\,1.6 \\
	B139 & FRB 20251013A & 60961.05698 & 20.9 & 173.5\,$\pm$\,0.4 \\
	B140 & FRB 20251013B & 60961.63358 & 10.2 & 174.3\,$\pm$\,0.4 \\
	B141 & FRB 20251014C & 60962.09928 & 12.4 & 174.3\,$\pm$\,0.4 \\
	B142 & FRB 20251014D & 60962.54879 & 15.8 & 173.5\,$\pm$\,0.4 \\
	B143 & FRB 20251015B & 60963.06541 & 22.6 & 173.5\,$\pm$\,0.4 \\
	B144 & FRB 20251023A & 60971.60560 & 11.1 & 173.5\,$\pm$\,0.4 \\
	B145 & FRB 20251027A & 60975.57843 & 8.6 & 175.1\,$\pm$\,0.4 \\
	B146 & FRB 20251031A & 60979.50794 & 8.7 & 173.5\,$\pm$\,0.4 \\
	B147 & FRB 20251107A & 60986.93720 & 10.0 & 177.1\,$\pm$\,0.8 \\
	B148 & FRB 20251117B & 60996.99944 & 10.8 & 173.5\,$\pm$\,0.4 \\
	B149 & FRB 20251129D & 61008.42769 & 11.1 & 174.3\,$\pm$\,0.4 \\
	B150 & FRB 20251202C & 61011.96716 & 14.1 & 175.5\,$\pm$\,0.8 \\
	B151 & FRB 20251203B & 61012.99213 & 9.7 & 175.5\,$\pm$\,0.8 \\
	B152 & FRB 20251213B & 61022.89104 & 12.5 & 175.5\,$\pm$\,0.8 \\
	B153 & FRB 20251217A & 61026.27752 & 8.2 & 174.3\,$\pm$\,0.4 \\
	B154 & FRB 20251219B & 61028.94067 & 13.4 & 173.5\,$\pm$\,0.4 \\
	B155 & FRB 20251230B & 61039.85693 & 16.0 & 173.9\,$\pm$\,0.8 \\
	\enddata
	\vspace{0.4em}
	\tablecomments{\hspace*{-2.44\tabcolsep}\parbox{0.82\textwidth}{List of radio bursts detected from FRB~20240209A between 2018~July~25~(MJD~58324) and 2025~December~31~(MJD~61040) with the \mbox{CHIME/FRB}~radio telescope.}}
	\hspace*{-2.46\tabcolsep}\parbox{0.82\textwidth}{\tablenotetext{a}{\,Transient Name Server~(TNS) name of each burst. The TNS naming scheme for FRB~bursts is described in~\citet{Yaron+2020} and~\citet{CHIME+2021b}.}}
	\vspace{-0.4em}
	\hspace*{-2.86\tabcolsep}\parbox{0.82\textwidth}{\tablenotetext{b}{\,Topocentric time of arrival~(ToA) of each burst, referenced to 400\,MHz. The ToAs were corrected for dispersion using a dispersion constant of $\mathcal{D}$\,$=$\,1/(2.41\,$\times$\,10$^{\mathrm{-4}}$)\,MHz$^{\mathrm{2}}$\,pc$^{\mathrm{-1}}$\,cm$^{\mathrm{3}}$\,s and the dispersion measure~(DM) listed for each burst. All burst ToAs are reported in the UTC time standard.}}
	\vspace{-0.4em}
	\hspace*{-2.86\tabcolsep}\parbox{0.82\textwidth}{\tablenotetext{c}{\,Detection signal-to-noise ratio~(S/N) of each burst, determined by the \mbox{CHIME/FRB} real-time FRB search pipeline.}}
	\vspace{-0.4em}
	\hspace*{-2.86\tabcolsep}\parbox{0.82\textwidth}{\tablenotetext{d}{\,Detection~DM of each burst, as measured by the \mbox{CHIME/FRB} real-time FRB~search pipeline. The quoted uncertainty reflects the coarse-grained DM~resolution from the \texttt{bonsai} real-time tree-dedispersion algorithm~\citep{CHIME+2018}.}}
	\label{table:frb20240209a_burst_log}
\end{deluxetable*}


\clearpage

\begingroup
\makeatletter
\let\@LN@col\relax
\makeatother
\nolinenumbers

\noindent\makebox[\textwidth][l]{
		\hspace*{-1.1in}
		\begin{minipage}{\dimexpr\textwidth+2.03in\relax}
			\startlongtable
			\begin{deluxetable*}{ccccccc}
				\tabletypesize{\small}
				\setlength{\tabcolsep}{9pt}
				\tablewidth{0pt}
				\tablecaption{Summary of FRB~Periodicity Search Results}
				\tablehead{\\[-15pt]\colhead{Source Name\,\tablenotemark{a}} & \colhead{Number of} & \colhead{Activity} & \colhead{Statistical} & \colhead{Detection} & \colhead{Active Cycles / Total Cycles\,\tablenotemark{f}} & \colhead{Trial Period} \\[-5pt]
					\colhead{} & \colhead{Bursts\rlap{\,\tablenotemark{b}}} & \colhead{Period\rlap{\,\tablenotemark{c}}} & \colhead{Significance\rlap{\,\tablenotemark{d}}} & \colhead{Statistic\rlap{\,\tablenotemark{e}}} & \colhead{} & \colhead{Range\rlap{\,\tablenotemark{g}}} \\
					\colhead{} & \colhead{($N$)} & \colhead{($P$)} & \colhead{} & \colhead{} & \colhead{($N_{\mathrm{cycles}}^{\mathrm{active}}$ / $N_{\mathrm{cycles}}^{\mathrm{total}}$)} & \colhead{[$P_{\mathrm{min}}$, $P_{\mathrm{max}}$]} \\
					\colhead{} & \colhead{} & \colhead{(days)} & \colhead{} & \colhead{} & \colhead{} & \colhead{[days, days]}}
				\startdata
				FRB~20180814A & 31 & \nodata & \nodata & \nodata & \nodata & [10.0, 250.3] \\
				FRB~20180916B & 142 & 16.335\,$\pm$\,0.008~[1]\,\tablenotemark{h} & 13.8$\sigma$~[1]\,\tablenotemark{i} & $Z_{1}^{2}$~[1] & 59 / 115.0 & [10.0, 582.7] \\
				FRB~20181119A & 16 & \nodata & \nodata & \nodata & \nodata & [10.0, 538.5] \\
				FRB~20181128A & 12 & \nodata & \nodata & \nodata & \nodata & [10.0, 570.8] \\
				FRB~20190117A & 33 & \nodata & \nodata & \nodata & \nodata & [10.0, 566.1] \\
				FRB~20190208A & 39 & 28.46\,$\pm$\,0.08 & 3.1$\sigma$ & $Z_{1}^{2}$ & 16 / 66.0 & [10.0, 525.2] \\
				FRB~20190212A & 13 & \nodata & \nodata & \nodata & \nodata & [10.0, 387.9] \\
				FRB~20190303A & 47 & \nodata & \nodata & \nodata & \nodata & [10.0, 551.8] \\
				FRB~20190417A & 19 & \nodata & \nodata & \nodata & \nodata & [10.0, 395.3] \\
				FRB~20190804E & 16 & \nodata & \nodata & \nodata & \nodata & [10.0, 328.8] \\
				FRB~20190907A & 17 & \nodata & \nodata & \nodata & \nodata & [10.0, 395.6] \\
				FRB~20191106C & 18 & \nodata & \nodata & \nodata & \nodata & [10.0, 335.7] \\
				FRB~20200926A & 13 & 86.7\,$\pm$\,0.5 & 4.0$\sigma$ & \mbox{$H$-test} & 4 / 21.7 & [10.0, 347.4] \\
				FRB~20200929C & 22 & \nodata & \nodata & \nodata & \nodata & [10.0, 311.8] \\
				FRB~20201124A & 37 & \nodata & \nodata & \nodata & \nodata & [10.0, 100.4] \\		
				FRB~20201130A & 12 & \nodata & \nodata & \nodata & \nodata & [10.0, 42.5] \\
				FRB~20220505A & 18 & \nodata & \nodata & \nodata & \nodata & [10.0, 164.9] \\
				FRB~20220529A & 31 & 36.4\,$\pm$\,0.5 & 3.5$\sigma$ & LS & 6 / 51.6 & [10.0, 128.6] \\
				FRB~20220912A & 476 & \nodata & \nodata & \nodata & \nodata & [10.0, 122.7] \\
				FRB~20240209A & 155 & 170.4\,$\pm$\,4.7\,\tablenotemark{j} & 5.3$\sigma$ & LS & 6 / 15.9 & [10.0, 429.2] \\
				\enddata
				\vspace*{0.4cm}
				\hspace*{-0.59cm}
				\begin{minipage}{0.86\textwidth}
					\tablecomments{Results of our search for periodic activity in the \mbox{CHIME/FRB} burst arrival times from FRB~20240209A and 19~repeating~FRB sources in \mbox{CHIME/FRB} Catalog~2. For each source with a candidate activity period identified by our search pipeline, we list the activity period, along with its trials-corrected statistical significance and the detection statistic that yielded the highest significance at the adopted period. We also report the number of active cycles (i.e.,~those containing at least one burst) and the total number of cycles spanned by the observations, both inferred from the listed activity period. In addition, we provide the range of trial periods searched for each source, including those for which no candidate periodicity was identified. We do not list an activity period for sources that did not satisfy the vetting criteria described in Section~\ref{sec:frb_periodicity_search_pipeline_overview}. References: [1]~\citet{Pearlman+2026a}.}
					\tablenotetext{a}{\,Transient Name Server~(TNS) name of the repeating~FRB source. The TNS naming scheme for FRB~sources is described in~\citet{Yaron+2020} and~\citet{CHIME+2021b}.}
					\vspace{-0.4em}
					\tablenotetext{b}{\,Number of \mbox{CHIME/FRB} burst~arrival times used in the periodicity search.}
					\vspace{-0.4em}
					\tablenotetext{c}{\,Preferred activity period.}
					\vspace{-0.4em}
					\tablenotetext{d}{\,Trials-corrected statistical significance of the activity period.}
					\vspace{-0.4em}
					\tablenotetext{e}{\,Detection statistic used to measure the activity period and its trials-corrected statistical significance: \mbox{Lomb--Scargle}~(LS), Rayleigh~($Z_1^2$), or \mbox{$H$-test}.}
					\vspace{-0.4em}
					\tablenotetext{f}{\,Number of cycles containing at least one burst and total number of cycles spanned by the observations. Both quantities are inferred from the listed activity period.}
					\vspace{-0.4em}
					\tablenotetext{g}{\,Range of trial periods searched for periodicity.}
					\vspace{-0.4em}
					\tablenotetext{h}{\,We quote the refined activity period for FRB~20180916B reported in~\citet{Pearlman+2026a}. An updated measurement of its activity period derivative ($\dot{P}$\,$=$\,$-$1.3$^{\mathrm{+4.0}}_{\mathrm{-4.0}}$\,$\times$\,10$^{\mathrm{-5}}$\,day\,day$^{\mathrm{-1}}$) is also reported in~\citet{Pearlman+2026a}.}
					\vspace{-0.4em}
					\tablenotetext{i}{\,See~\citet{CHIME+2020b} for an additional measurement of the statistical significance of FRB~20180916B's \mbox{16.3-day} activity period, based on a smaller number of burst arrival times spanning fewer total cycles.}
					\vspace{-0.4em}
					\tablenotetext{j}{\,FRB~20240209A also exhibits very tentative evidence for an activity period derivative~($\dot{P}$\,$=$\,0.20$^{\mathrm{+0.04}}_{\mathrm{-0.04}}$\,day\,day$^{\mathrm{-1}}$; see Section~\ref{sec:frb20240209a_results}).}
					\label{table:frb_periodicity_search_summary}
				\end{minipage}
			\end{deluxetable*}
		\end{minipage}
}

\linenumbers
\endgroup

\clearpage


\bibliography{references}
\bibliographystyle{aasjournalv7}

\end{document}